\documentclass{jfm}
\usepackage{graphicx}
\usepackage{epstopdf, epsfig}
\usepackage{amsmath}
\usepackage{color}
\usepackage{rotating}

\begin{document}

\newtheorem{lemma}{Lemma}
\newtheorem{corollary}{Corollary}

\shorttitle{Turbulent channel flow of a dense binary mixture of rigid particles} 
\shortauthor{I. Lashgari, F. Picano, P. Costa , W.-P Breugem and L. Brandt} 

\title{Turbulent channel flow of a dense binary mixture of rigid particles}

\author
 {
 Iman Lashgari\aff{1}
  \corresp{\email{imanl@mech.kth.se}},
  Francesco Picano\aff{2},
  Pedro Costa\aff{3}, \\
   Wim-Paul Breugem\aff{3}
  \and 
  Luca Brandt\aff{1}
  }

\affiliation
{
\aff{1}
 Linn\'e FLOW Centre and SeRC (Swedish e-Science Research Centre),KTH Mechanics, SE-100 44 Stockholm, Sweden
\aff{2}
Department of Industrial Engineering, University of Padova, Padova, Italy
\aff{3}
Laboratory for Aero and Hydrodynamics, Delft University of Technology, Delft, The Netherlands
}

\maketitle

\begin{abstract}
{We study turbulent channel flow of a binary mixture of finite-size neutrally-buoyant rigid particles by means of interface-resolved direct numerical simulations. We fix the bulk Reynolds number and total solid volume fraction, $Re_b = 5600$ and $\Phi=20\%$, and vary the relative fraction of small and large particles. The binary mixture consists of particles of two different sizes, $2h/d_l=20$ and $2h/d_s=30$ where $h$ is the half channel height and $d_l$ and $d_s$ the diameter of the large and small particles. While the particulate flow statistics exhibit a significant alteration of the mean velocity profile and turbulent fluctuations with respect to the unladen flow, the differences between the mono-disperse and bi-disperse cases are small. However, we observe a clear segregation of small particles at the wall in binary mixtures, which affects the dynamics of the near wall region and thus the overall drag. This results in a higher drag in suspensions with a larger amount of large particles. As regards bi-disperse effects on the particle dynamics, a non-monotonic variation of the particle dispersion in the spanwise (homogeneous) direction is observed when increasing the percentage of small/large particles. Finally, we note that particles of the same size tend to cluster more at contact whereas the dynamics of the large particles gives highest collision kernels due to a higher approaching speed. }  

\end{abstract}

\section{Introduction}

The turbulent flow of particle suspensions is a complicated problem in fluid mechanics because both turbulence and the behaviour of particles in suspensions are not completely understood and the combination of the two raises new significant challenges \cite[][]{Balachandar10,Prosperetti15}. Despite its complexity, this flow serves many natural and practical applications from pyroclastic and sedimentation flows to fluidised beds, hopper dredger and slurry transports \cite[][]{Eckstein77}. In this work, we employ direct numerical simulations to study turbulent particulate channel flow where particles have two different sizes, i.e. binary mixtures. The fully-resolved simulation of the dispersion of thousands of rigid particles in a turbulent flow enables us to gain new understanding of the microphysics of the problem, developing theories and predicting the complex behaviour of the system, \cite[][]{Lucci10,Campbell90}.

The interactions between the particles and fluid depend strongly on the characteristics of both the disperse and the continuous phases. If the volume fraction of the particles is larger than a certain threshold, e.g. $10^{-3}$ according to the analysis in \cite{Elghobashi94}, full interactions between the two phases take place, leading to the so-called four-way coupling regime. This condition is typical in many industrial processes such as crystallisation and fluidised beds. To study numerically the fully-coupled regime of a suspension the classical point-particle models, based on Maxey and Riley's model \cite[][]{Maxey83}, are no longer valid and direct numerical simulations resolving the flow around each individual particle should be employed. The same holds when the particle size is larger than the smallest scale in the flow, even for dilute suspensions.

As regards suspensions of finite size particles, most of the previous studies are conducted at zero or low inertia with main focus on the rheological aspects of the flow, in particular particle distribution, effective viscosity and normal stress differences \cite[see][]{Hampton97,Stickel05,Brown09,mor_ra09,Maxey11,Yeo13}. When inertia at the particle scale is large enough, the symmetry of the flow around the particles is broken and this affects the rheological behaviour of the suspension \cite[][]{Morrispof08,Picano13,Morris14}. Given the range of applications, suspensions flowing in the inertial regime have been studied long, starting from the seminal work of  \cite{Bagnold54} who performed experiments with suspensions of rigid particles between two cylindrical drums. Bagnold defines two regimes at low and high shear rates, macro-viscous and grain inertia, where the effective viscosity of the suspension, measured by the wall shear stress, varies linearly and quadratically with shear rate, respectively. Several decades later, \cite{Matas03} conducted experiments on suspensions of finite-size particles in a pipe and reported the critical threshold for the transition from the laminar to the turbulent regime as a function of the particle size and volume fraction. Most importantly, these authors observe that the critical threshold first decreases and then increases with the volume fraction of the dispersed phase when the particles are large enough with respect to the pipe diameter. The transition promotion is attributed to the disturbances induced by the particles as well as to the breakdown of the flow coherent structures \cite[see][]{Loisel13,Lashgari15}. The simulations in \cite{Lashgari14} aimed to reproduce the experiments in \cite{Matas03}; these allow the authors to introduce three different regimes: laminar, turbulent and inertial shear-thickening for channel flow laden with finite-size neutrally buoyant particles when varying the Reynolds number, $Re$, and the particle volume fraction, $\Phi$. These regimes are identified considering the different contribution of the viscous, Reynolds and particle stresses to the stress budget of the two phase flow. The regimes are characterized by different mechanisms governing the particle dynamics and present different dispersion features \cite[see][]{Lashgari16}.

The recent advancement of computational resources and the improvement of the numerical algorithms has allowed the scientific community to address turbulent flows laden with thousands of finite-size particles. In this respect, the early work of \cite{Tencate04} employ a lattice-Boltzmann scheme to simulate the motion of finite size particles in an isotropic turbulent flow. These authors show that the kinetic energy and energy dissipation at wavelengths close to the particle size increase. This finding has been confirmed experimentally by \cite{Bellani12} for both spherical and non-spherical particles. We use here the Immersed Boundary Method, which was first adopted by \cite{Uhlmann05} to simulate flow of rigid particles in suspension. \cite{Uhlmann08}
simulated thousands of finite-size particles at moderately high particle Reynolds number in a vertical turbulent channel flow mimicking the flow in fluidised bed. In this study, the appearance of large structures in the flow is related to the instabilities induced by the particles while an apparent segregation of the particles is not observed. Later, the same numerical approach has been employed to simulate flow in other configurations such as turbulent particulate flow in an open channel and sedimentation  \cite[][]{Kidanemariam13,Fornari16a}. 
 
As regards turbulent channel flow laden with finite size particles, \cite{Picano15} study the flow of neutrally-buoyant particles at volume fractions up to $20\%$. They report that the particles alter the near-wall dynamics by increasing the spacing of the streaks and reducing their velocity contrast. More importantly,  the log region in the mean fluid velocity profile changes considerably in the presence of the particles. Despite the monotonic increase in the overall drag with the particle concentration, these authors report a reduction of the turbulence activity at high volume fractions. Recently, \cite{Fornari16b} extended this first study considering dense suspensions with different particle to fluid density ratios, $R$, up to $1000$, yet neglecting settling. This study shows that the excluded volume effect, i.e.\ the particle volume fraction, determines the turbulence statistics, while the particle to fluid inertia has negligible effect when $R<10$. In addition, it is shown that strong particle shear-induced migration to the channel centreline occurs for the case of  $R=10$ and the particle dynamics decouple from the fluid at $R=1000$, but not vice-versa. 

Theoretical analysis of the multi-scale dynamics of the unladen turbulent channel flow in the framework of the so-called "Law of the wall" results in a reasonable prediction of the wall shear stress as the function of bulk Reynolds number;  a commonly used approximate relation for the friction Reynolds number is $Re_\tau= 0.09(Re_b)^{0.88}$ \cite[][]{Pope00}. Being able to extend this law to the turbulent particulate channel flow is desirable to avoid massive numerical and experimental investigations needed to understand the characteristics of the flow in different conditions. Recently \cite{Costa16} proposed a new theoretical framework to estimate the bulk behaviour, overall drag and mean velocity profile of turbulent channel flow laden with mono-disperse neutrally-buoyant rigid particles. These authors show that the channel can be divided into a near wall region where the particle form an evident layer and the rest of the channel where the particle distribution is homogenised. The thickness of the near-wall region is modelled as a function of particle size and volume fraction. As we will show here the proposed model contributes directly to the understanding of the behaviour of the wall friction in a bi-disperse turbulent suspension as well, when properly accounting for the particle size distribution.

 In almost all studies mentioned above, either in the laminar or turbulence regime, the particle phase is mono-disperse, i.e.\ particles have the same size. This assumption is not valid in many applications where suspensions contain a wide range of particle sizes. Poly-dispersity introduces additional non-uniformity in the flow that may affect the particle dynamics as well as turbulent flow dynamics \cite[][]{Marchioro00}. Hence, the aim of the present study is to assess similarities and differences between mono-disperse and bi-disperse turbulent suspensions. As a first step, we therefore consider the simpler case of binary mixtures and perform simulations of turbulent channel flow of bi-disperse particle suspensions at fixed bulk Reynolds number, $Re_b=5600$, and total volume fraction, $\Phi=0.2$. { At this concentration, the effect of the particle-induced stress on the momentum transfer across the channel becomes important as shown by \cite{Picano15}}. We vary the ratio between the relative volume fraction of small and large particles.  We focus on the analysis of the bulk behaviour of the suspension flow and on the particle dynamics providing local concentration, dispersion and collision to compare with the studies by \cite{Picano15,Lashgari16}. This work can be seen as a step forward in exploring the wide parameters space of particulate flows in the journey towards the more realistic simulations of suspensions \cite[][]{Prosperetti15}.      

The paper is organised as follows. We discuss the governing equations and numerical method in $\S2$, whereas  the flow configuration and simulation specifications are reported in $\S3$. The results of the simulations are presented in $\S4$ with conclusions and final remarks discussed in $\S5$. Results pertaining binary mixtures in the laminar regime are reported in the Appendix and used as a reference when discussing the results for the turbulent flow.

\section{Governing equations and numerical method}

We study a turbulent channel flow laden with a binary suspension of finite size particles where the carrier phase is a Newtonian and incompressible fluid
and the solid phase constituted by rigid, neutrally-buoyant spheres. The fluid flow is governed by Navier-Stokes and continuity equations,
\begin{align}
\rho (\frac{\partial \textbf{u}}{\partial t} + \textbf{u} \cdot \nabla \textbf{u}) = -\nabla P + \mu \nabla^2 \textbf{u} + \rho \textbf{f}, \\ \nonumber
\nabla \cdot \textbf{u} = 0,
\label{eq:NS}  
\end{align}
where $\mu$ and $P$  indicate the fluid dynamic viscosity and pressure  and $\rho$  is the density of both fluid and particles. The coordinate system and velocity components are denoted by $\textbf{X}=(x,y,z)$ and $\textbf{u}=(u,v,w)$ corresponding to streamwise, wall normal and spanwise directions. A localized force $\textbf{f}$ is added on the right hand side of the Navier-Stokes equation to treat the presence of finite-size particles by means of an Immersed Boundary Method. The motion of the particles is governed by the Newton-Euler equations,
\begin{align}
m^p \frac{ d \textbf{U}_c^{p}}{dt} = \oint_{\partial {V}_p}  [ -P\textbf{I} + \mu (\nabla \textbf{u} + \nabla \textbf{u}^T ) ] \cdot  \textbf{n} dS+ \textbf{F}_c , \\ \nonumber
I^p \frac{ d \pmb{\Omega}_c^{p}}{dt} = \oint_{\partial {V}_p}  \textbf{r} \times \big{\{} [ -P\textbf{I} + \mu (\nabla \textbf{u} + \nabla \textbf{u}^T ) ] \cdot  \textbf{n}  \big{\}} dS + \textbf{T}_c, 
\label{eq:NE}  
\end{align}
where the mass, moment inertia, centroid velocity and angular velocity of the particle $p$ are denoted by $m^p$ and $I^p$, $\textbf{U}_c^p$ and $\pmb{\Omega}_c^p$, respectively. The surface of the particles and unit normal vector are denoted by $\partial {V}_p$ and  $\textbf{n}$, whereas the vector connecting the centre to the surface of the particles is indicated by $\textbf{r}$.  The first term on the right hand side of these equations represents the net force/moment on the particle $p$ resulting from the surrounding flow. The second term, $\textbf{F}_c$ and $\textbf{T}_c$, represent the force and torque resulting from short-range interactions, lubrication and collision. The interface condition is introduced to enforce the fluid velocity at each point on the particle surface to be equal to the particle velocity at that point,  $\textbf{u}(\textbf{X})= \textbf{U}^p (\textbf{X}) = \textbf{U}_c^p + \pmb{\Omega}_c^p \times \textbf{r}$. The Immersed Boundary Method with direct forcing developed by \cite{Uhlmann05} and modified by \cite{Breugem12} is employed to integrate the particle motion  and  satisfy the interface condition by the forcing  $\textbf{f}$ in the vicinity of each particle surface.

The fluid flow dynamics is solved discretising the incompressible Navier-Stokes equations with a second-order finite difference method on a staggered grid. 
The solver is based on the discrete forcing method to simulate neutrally buoyant particles with second-order spatial accuracy \cite[][]{Mittal05,Breugem12}. Two sets of grid points are considered: an Eulerian fixed and equispaced three-dimensional mesh and a set of Lagrangian points uniformly distributed on the surface of each particle. The Eulerian and Lagrangian grid points communicate to compute the Immersed Boundary forcing and ensure the no-slip and no-penetration boundary conditions on the surface of the particles. The IB force is then applied on both fluid and solid phases to evolve velocities and positions.      

When the gap width between two particles (or particle-wall) becomes less than a threshold value, $\epsilon$, the IBM underestimates the actual lubrication force. Therefore, a resolution dependent lubrication correction  is included at small $\epsilon$ as a function of the gap width. This correction is kept constant below a second threshold to represent the surface roughness. When $\epsilon \le 0$, a collision takes place; in this case, the lubrication correction is turned off and a collision force is activated \cite[see the Appendix of][for more details]{Lambert13}. The collision force is computed based on the particle relative velocity and overlap. We use here the soft-sphere collision model described in the recent work by \cite{Costa15} where a mass-spring-damper system in the directions normal and tangential to the contact line between the two overlapping spheres (or sphere-wall) govern the dynamics of the collision. Since the collision time is generally much smaller than the viscous relaxation time, even for a wet collision, the collision model allows us to stretch the collision time artificially so as to avoid limiting restrictions of the numerical time step. The accuracy of the collision model has been tested against several benchmark cases in \cite{Costa15}.

\section{Computational setup}

\begin{table*}
  \begin{center}
    \setlength\tabcolsep{4ex}
  \begin{tabular*}{\textwidth}{*{6}{c}}
    \hline
    Case name               & $0-100$ & $25-75$ & $50-50$ & $75-25$ & $100-0$ \\ \hline
    $\Phi_s $       &   0.2    &  0.15  &   0.1    &   0.05  &   0        \\ \hline  
    $\Phi_l $        &     0      &  0.05  &   0.1    &   0.15  &   0.2      \\ \hline
    $np_s $         & 46140 &34807& 23205&11602  &   0        \\ \hline
    $np_l $          & 0          & 3438  & 6875  & 10313 &  13751    \\ \hline
  \end{tabular*}
  \caption{Parameters of the five simulations of turbulent flow of bi-disperse suspensions; $\Phi_s$: volume fraction of small particles, $\Phi_l$: volume fraction of large particles, $np_s$: number of finite-size small particles and $np_l$: number of finite-size large particles. The solid volume fraction  is fixed to $\Phi= 20\%$ for all cases. }
  \label{tab:tab1}
\end{center}
\end{table*}

We simulate suspensions of neutrally-buoyant finite-size particles in a plane channel with periodic boundary conditions imposed in the streamwise and spanwise directions.  The computational domain has size $6h\times2h\times3h$ in the streamwise, wall-normal and spanwise directions where $h$ is the half channel height. We simulate the flow at bulk Reynolds number,  $Re={2h\,U_b}/{\nu}=5600$, where $U_b$ is the bulk velocity of the entire mixture and $\nu$ the fluid kinematic viscosity. 
The simulations are performed forcing the bulk velocity to $U_b=1$.  
This corresponds to a friction Reynolds number, $Re_{\tau}= 180$ in the unladen case.
We use a resolution of $1440 \times 480 \times 720$ grid points  in the streamwise, wall-normal and spanwise directions. For all the particulate cases this gives a  grid spacing of $\Delta x^+ = \Delta y^+ =\Delta z^+ \le 1 $, a value below the criteria to ensure fully-resolved simulation of the turbulent flow in a channel. The choice of the number of grid points is determined by the minimum resolution required  to resolve the flow around the smallest particle in the domain.  

The solid phase consists of spherical particles of two different sizes, $2h/d_s=30$ and $2h/d_l=20$, where $d_s$ and $d_l$ are the diameter of the small and the large particles. The two type of particles considered have therefore a ratio of diameters of 1.5, which corresponds to larger particles with volume $3.375$ times that of the small particles. These values of the particle diameter correspond to $16$ and $24$ Eulerian grid points, respectively. The communication between the solid surface and the surrounding flows occurs via $746$ and $1721$ Lagrangian grid points located on the particle surface. The number of Lagrangian grid points provides a similar spacing on the surface of the particles as the one of the Eulerian points, which contributes to an accurate exchange of the forces between the two phases. The total particle volume fraction is kept constant to $\Phi=0.2$, while the ratio between the volume fraction of the small to large particles is changed in the different simulations performed. The details of the five simulations of a particle-laden flow presented in the present work are reported in table \ref{tab:tab1};  the case names reflect the percentage of the volume fraction occupied by large and small particles. The two cases $0-100$  and $100-0$ correspond therefore to  mono-disperse suspensions of only small and large particles, respectively.  

\begin{figure}
\begin{center}
   \includegraphics[width=10cm]{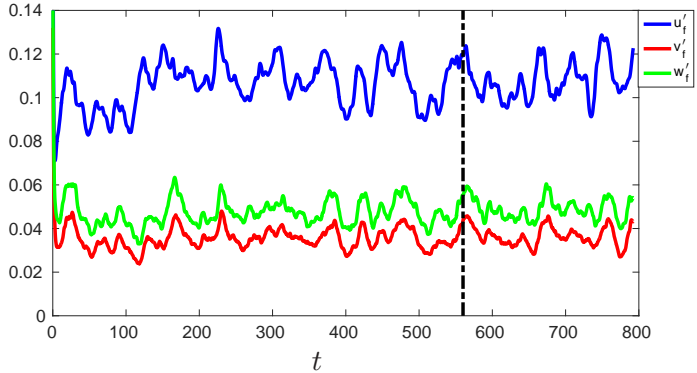}
    \put(-150,-5){{\large $t$}} 
   \caption{Time history of the fluid velocity fluctuations for the case $0-100$. The velocity fluctuations are scaled by $U_b$ whereas time is shown in bulk flow units, $2h/U_b$. }
   \label{fig:uvw}
\end{center}
\end{figure}

The simulations are initialised with a semi-organised (lattice-like) arrangement of the particles in the entire domain together with a high amplitude localised disturbance in the form of two counter-rotating streamwise vortices \cite[][]{Henningson91}. These vortices efficiently trigger transition to turbulence in the domain and mix the small and large particles.  Each simulation is run using 480 cores for about 4 weeks. The statistics are computed after the initial transient phase {using 100 snapshots over approximately 230 bulk flow time units, $2h/U_b$. We display in figure~\ref{fig:uvw}, the time history of the box-averaged fluid velocity fluctuations for the case $0-100$. Velocity fluctuations are scaled by $U_b$ whereas time is shown in bulk-flow units, $2h/U_b$. The statistics are calculated using the data after the initial transients, indicated by the vertical dashed line.} Convergence tests are carried out by comparing the  statistics calculated with half of the number of samples.

\section{Results}
  
\begin{figure}
\begin{center}
   \includegraphics[trim = 0mm 0mm 0mm 0mm, width=9cm]{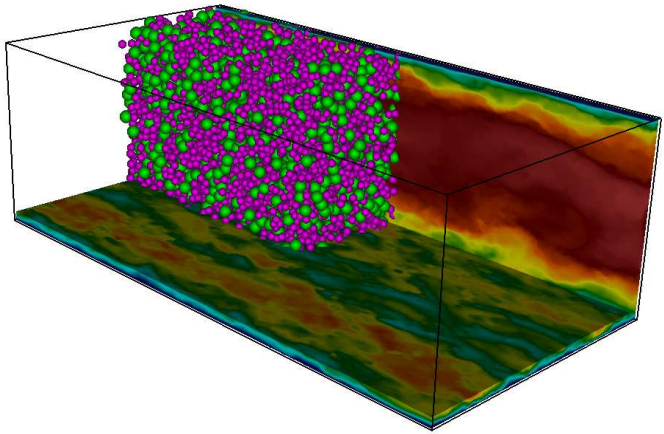}\\
   \caption{Instantaneous flow visualisation for a binary mixture with equal volume fraction of small and large particles: case $50-50$. The contour plot shows the streanwise velocity of the mixture in a wall-normal and wall-parallel plane. The arrangement of the particles is shown only over $1/4^{th}$ of the domain for the sake of clarity. The ratio between the channel height and the diameter of green and pink particles are $20$ and $30$ respectively.}
   \label{fig:vis}
\end{center}
\end{figure}

First we show in figure \ref{fig:vis} a visualisation of the instantaneous flow for a binary mixture with equal volume fraction of small and large particles: case $50-50$.  We display the color contours  of the streamwise velocity of the mixture in a wall-normal and wall-parallel plane. The mixture velocity is obtained averaging the fluid and particle velocities in the domain. 
In the wall-parallel plane close to the bottom wall of the channel we observe the classical streaky structures, elongated in streamwise directions. These streaks are created by  the lift-up mechanism and serve as engine for sustaining the turbulence \cite[][]{Brandt14}. The presence of the particles alters the features of the streaks by increasing the spacing and decreasing the contrast between the region of high and low velocities, see also \cite{Picano15} for monodisperse suspensions. In the figure, we also display the particle position over $1/4^{th}$ of the domain for the sake of clarity. The distribution of both the small and the large particles is almost uniform in the middle of the channel. However, stronger segregation of small particles in the near-wall region is evident as we will show in more detail in the following. A similar behaviour is observed by visualising the flow from the other simulations (not shown here).

 \subsection{Mean velocity and particle distribution}
 
\begin{figure}
\begin{center}
   \includegraphics[width=5.5cm]{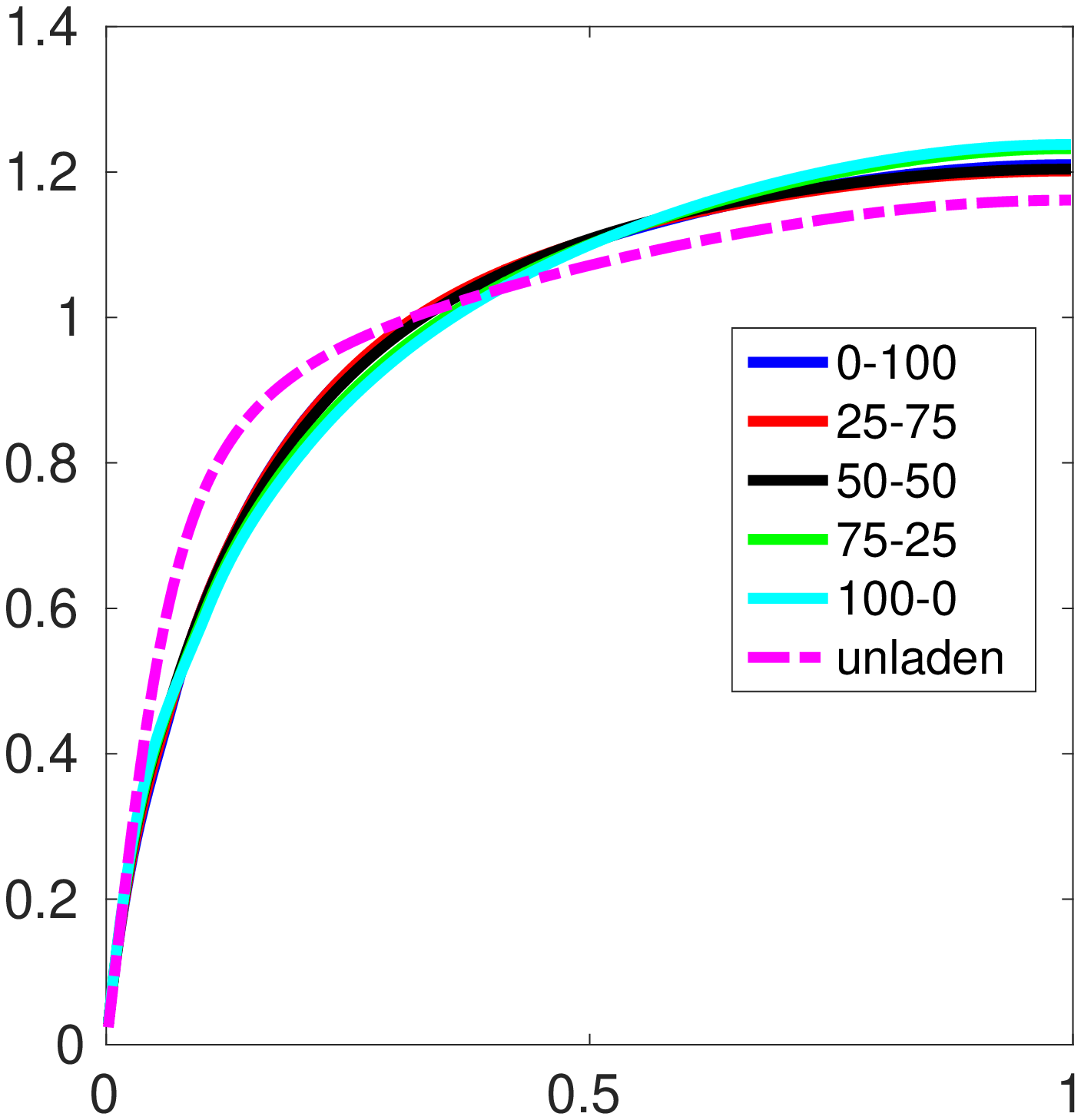}
   \put(-165,150){{\large $(a)$}} 
    \put(-162,80){{\large $U_f$}} 
    \put(-80,0){{\large $y/h$}} 
   \includegraphics[width=5.5cm]{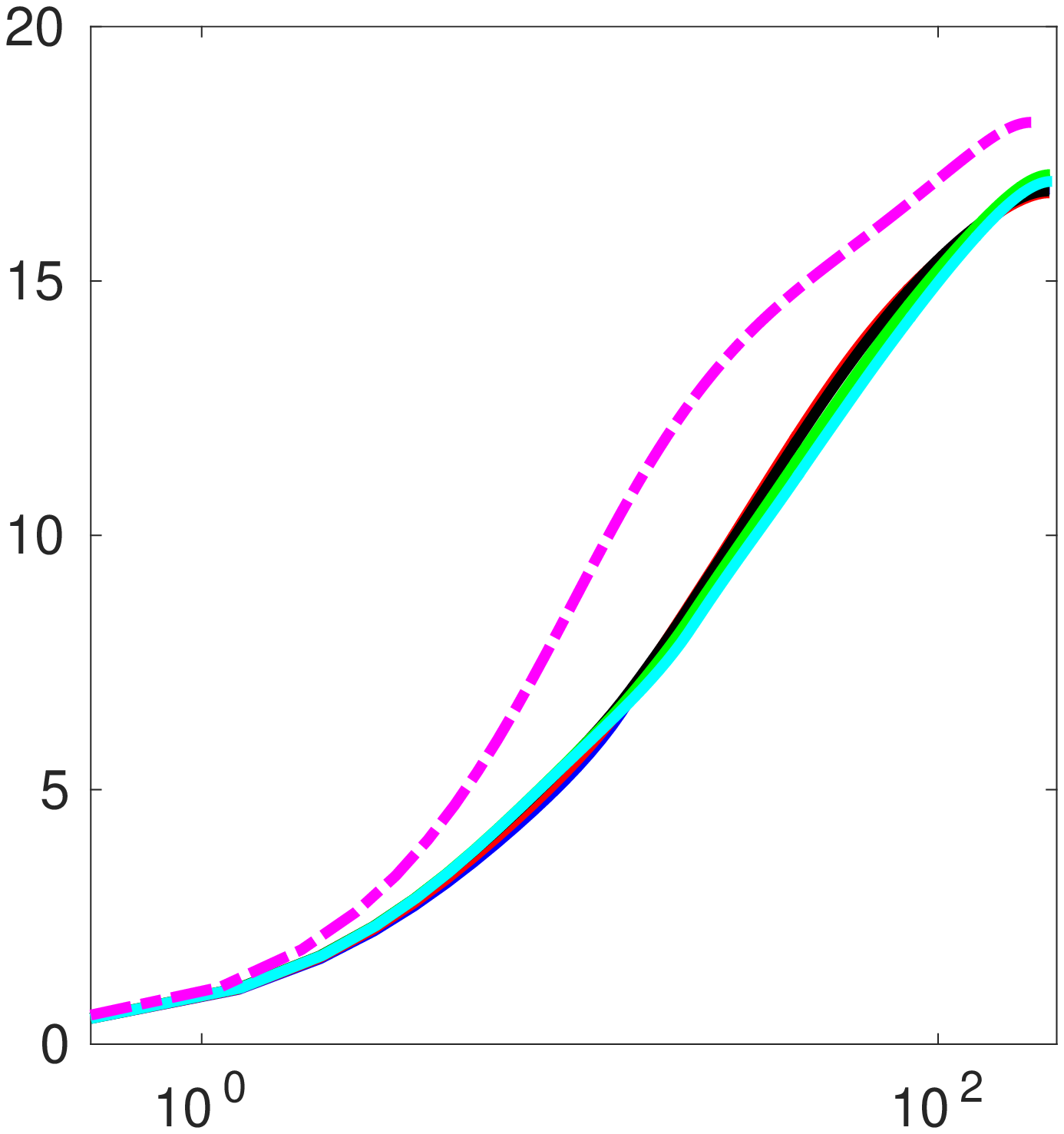}
    \put(-165,150){{\large $(b)$}} 
     \put(-162,80){{\large $U^+_f$}} 
    \put(-80,0){{\large $y^+$}} \\
    \includegraphics[width=5.5cm]{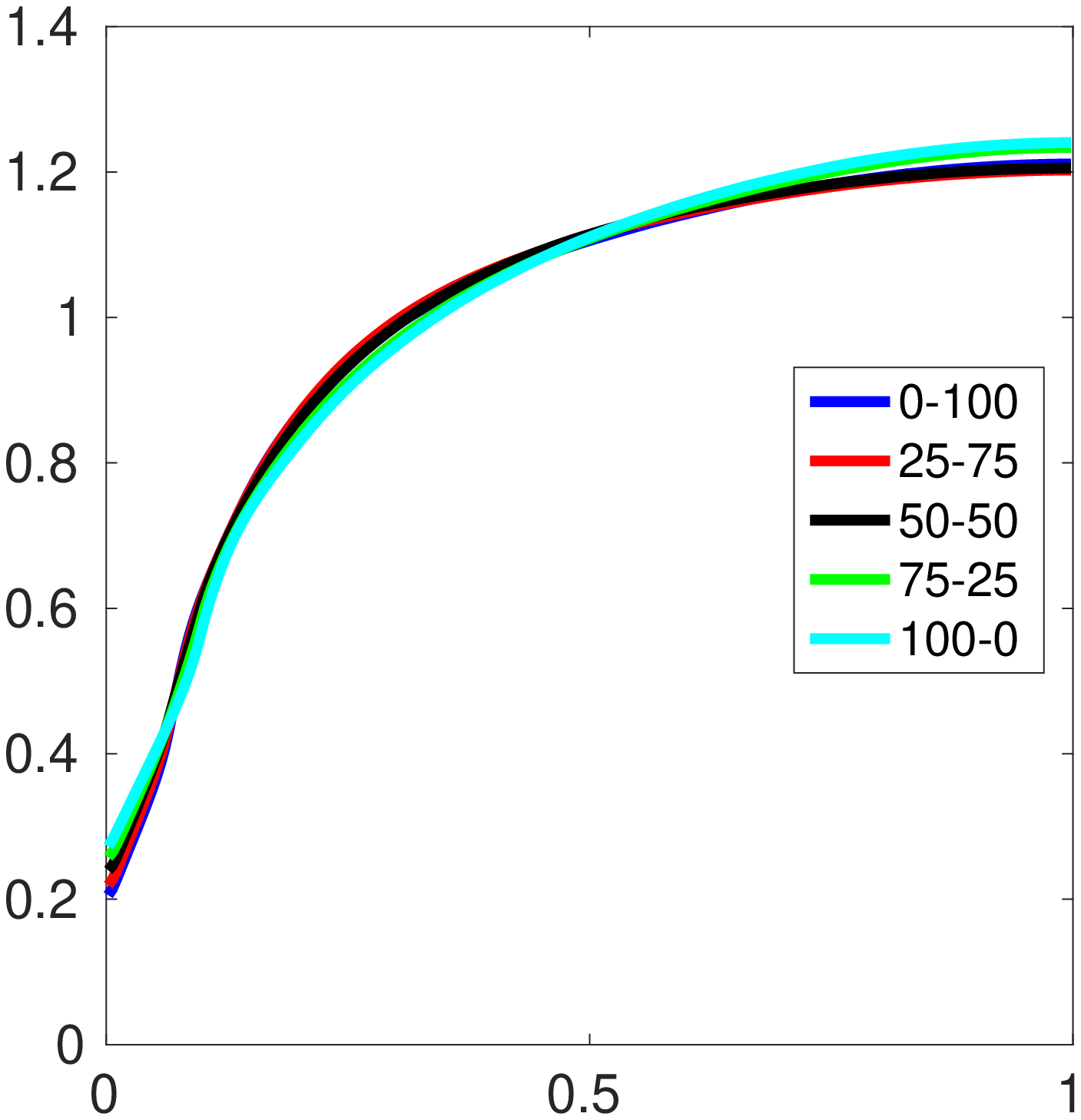}
    \put(-165,150){{\large $(c)$}} 
     \put(-162,80){{\large $U_{p}$}} 
    \put(-80,0){{\large $y/h$}} 
    \includegraphics[width=5.5cm]{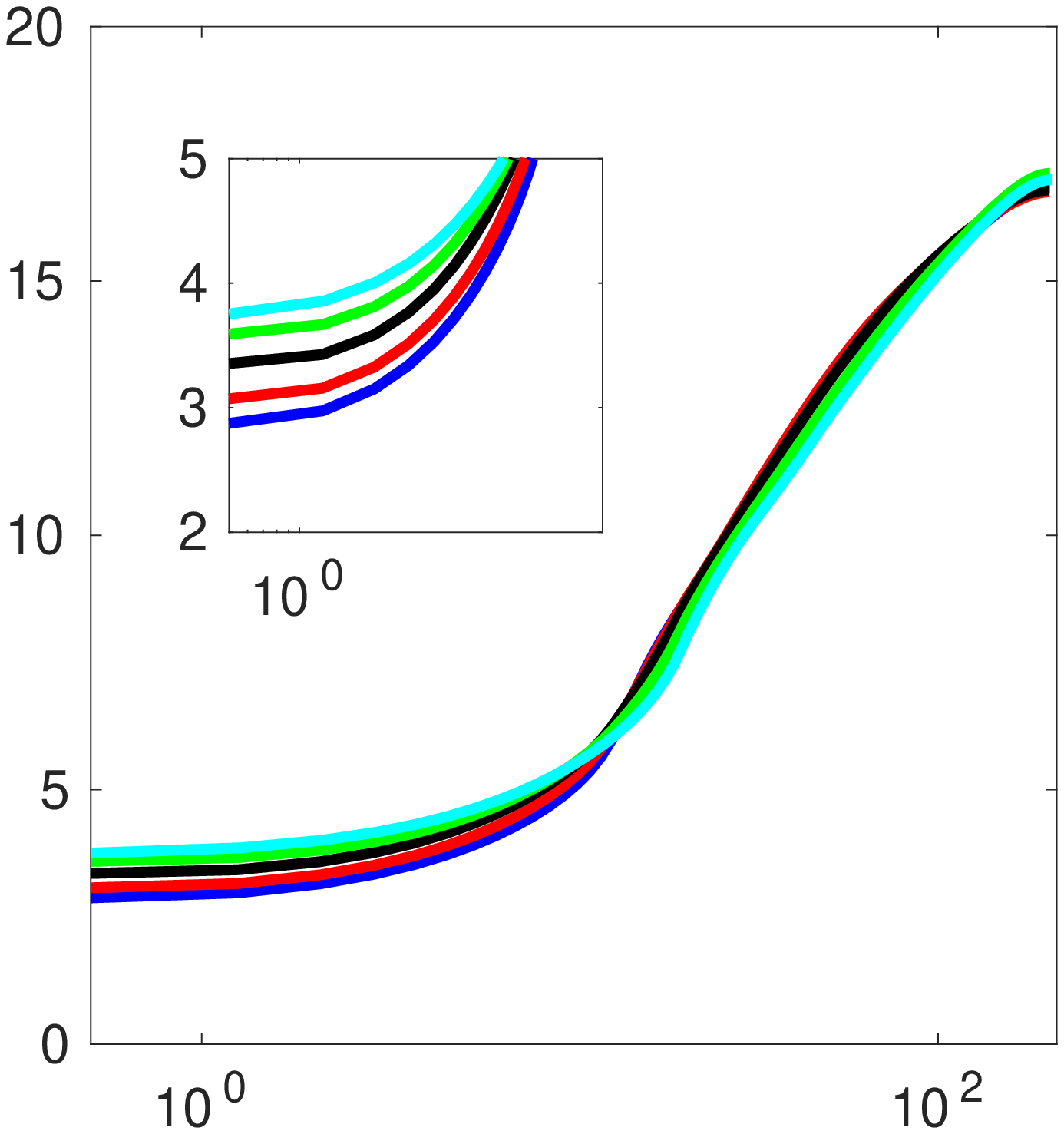}
    \put(-165,150){{\large $(d)$}} 
     \put(-162,80){{\large $U^+_p$}} 
    \put(-80,0){{\large $y^+$}} 
   \caption{Mean fluid and particle velocity profiles of particulate and unladen turbulent flow cases, a,c) in outer scaling b,d) in inner scaling.}
   \label{fig:mean}
\end{center}
\end{figure}

In  figure \ref{fig:mean}(a,b), we display the mean streamwise fluid velocity profile for the five cases under investigation. We also include the velocity profile of an unladen turbulent channel flow at the same bulk Reynolds number, $Re_b=5600$, indicated with dashed line. We recall that all the simulations are performed by enforcing a constant mass flux. The fluid statistics are obtained by averaging in the streamwise and spanwise directions over the points outside the particles and by temporal averaging. Although the mean velocity profile for the particle laden turbulent flows exhibit a strong modification with respect to the case of turbulent unladen flows, the difference between the particulate cases is not significant. In figure \ref{fig:mean}(a) we note that the mean velocity profile of the particulate flow is less blunt and more similar to a laminar flow with higher velocity toward the channel centre. This behavior is slightly accentuated in suspensions with higher concentration of large particles. This suggests that the turbulent activity is  reduced in the presence of the solid phase, the more so for larger particles. In figure \ref{fig:mean}(b) we present the mean velocity profile of particulate and unladen cases scaled in inner units, i.e. $U_f^+=U_f/u_{\tau}$ and $y^+=y u_{\tau}/\nu$ with $u_{\tau}=\sqrt{\tau_w/\rho}$ the friction velocity and $\tau_w$ the wall shear stress. The inner-scaled velocity profiles of the particulate cases are lower than those of the unladen flow indicating an increase in the wall-shear stress in the presence of particles. The log profile in the classical unladen turbulent flow reads $U^+= \frac{1}{\kappa} \mathrm{ln} y^+ + B$  where $\kappa$ is the Von $K \acute{a} rm \acute{a}n$ constant and $B$ is an additive coefficient \cite[][]{Pope00}. Here we obtain, similar to \cite{Picano15}, that the slope of the log region increases while the additive constant reduces significantly which results in an overall drag enhancement. No significant differences emerge among the different cases. We therefore conclude that the mean flow velocity of the turbulent channel flow is definitely controlled by the volume fraction of the solid phase, while the bi-dispersity of the suspensions plays a minor role.

Panel (c) of figure \ref{fig:mean} shows the mean velocity of the dispersed solid phase. 
Particle statistics are obtained by spatial averaging over the points inside the particles and then by temporal averaging. The data in the figure reveal, overall, a similar behaviour for the different particle-laden flows. Comparing with the data in figure \ref{fig:mean}(a)
one finds that slip between the fluid and particle velocities is evident close to the wall whereas this is almost zero in the core of the channel. 
The slip velocity is thus driving the dynamics near the wall, as further discussed below. {As shown in the inset of figure \ref{fig:mean}(d), the particle slip velocity at the wall increases monotonically when increasing the ratio between the volume fraction of large to small particles. This is attributed to the fact that larger particles forming the near-wall layer reach larger distances from the wall where they are exposed to higher flow velocity \cite[see][]{Picano15,Costa16}. }

\begin{figure}
\begin{center}
   \includegraphics[width=7.5cm]{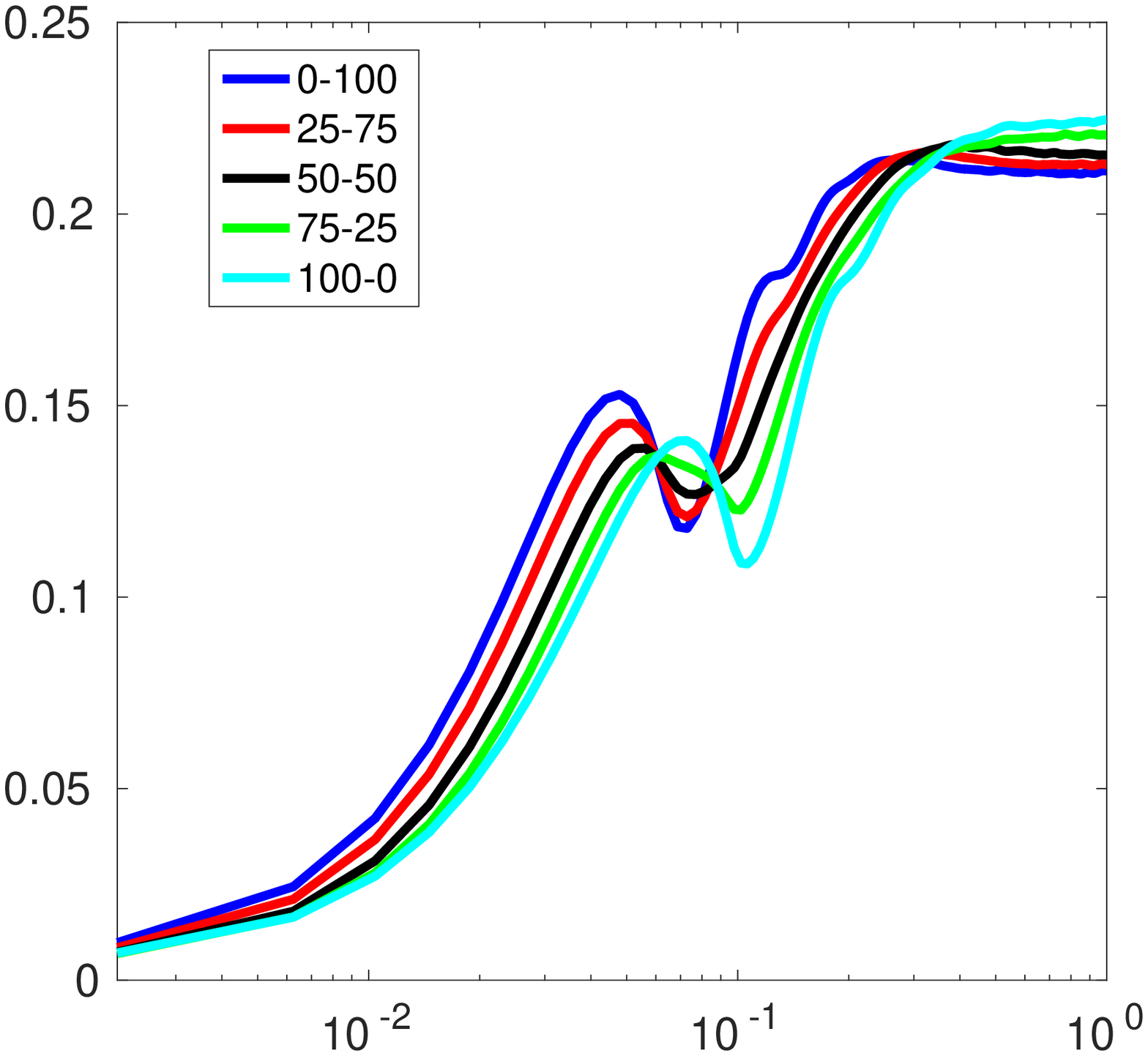}
      \put(-220,200){{\large $(a)$}} 
     \put(-220,100){{\large $\phi$}} 
     \put(-105,0){{\large $y/h$}} \\
   \includegraphics[width=5.5cm]{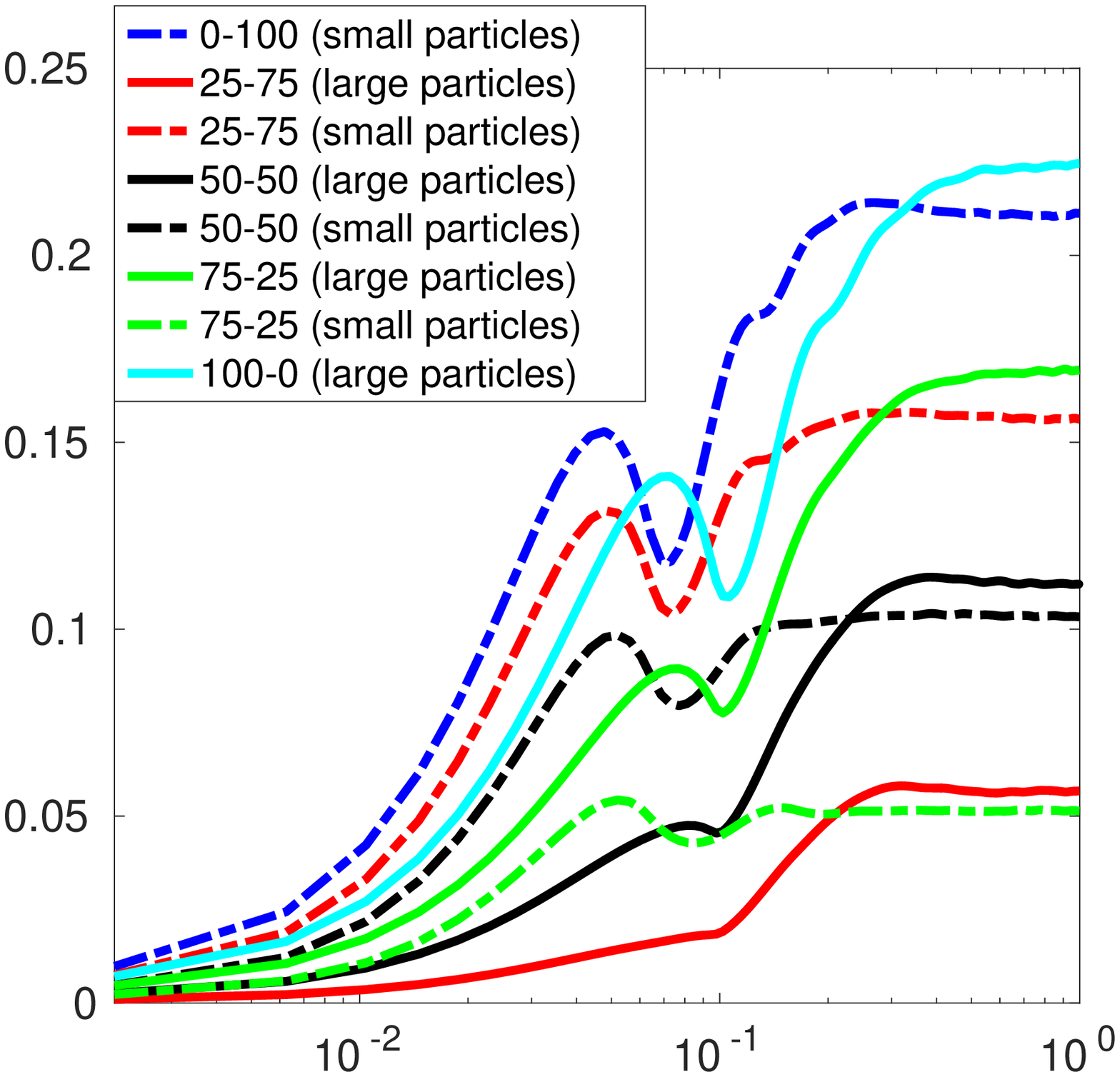}
      \put(-160,143){{\large $(b)$}} 
      \put(-160,80){{\large $\phi$}} 
      \put(-80,0){{\large $y/h$}}
   \includegraphics[width=5.5cm]{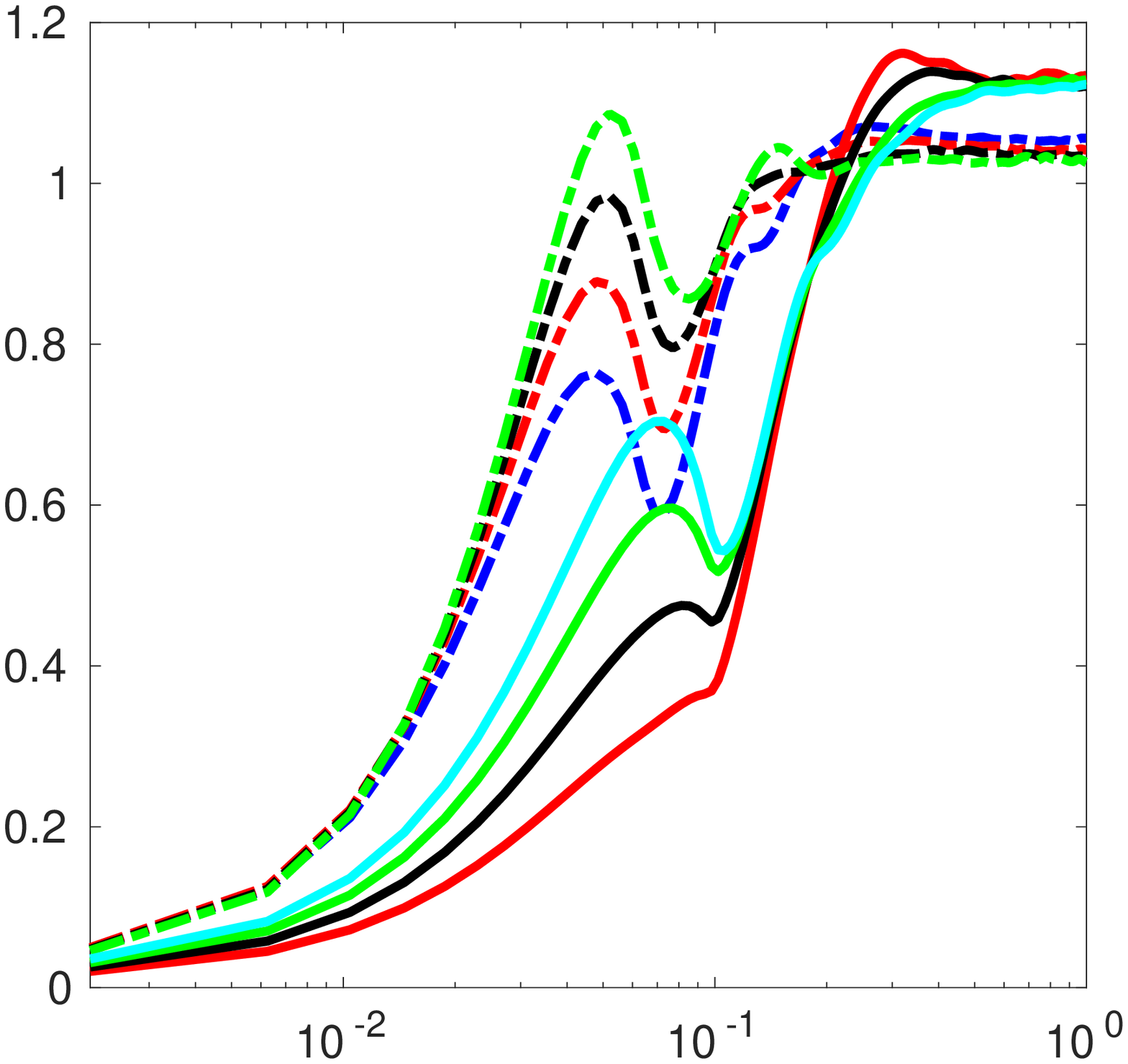}
      \put(-160,143){{\large $(c)$}} 
       \put(-10,80){{\large $\phi/\Phi$}} 
      \put(-80,0){{\large $y/h$}}
   \caption{Profile of local volume fraction for the five simulated suspensions, a) total volume fraction b) volume fraction of small and large particles separately c) volume fraction of small and large particles normalised by their bulk value.}
   \label{fig:localphi}
\end{center}
\end{figure}

 
{We employ a phase indicator function to calculate the local volume fraction profile across the channel. The indicator function assumes for each computational cell in the domain values of  $\Psi=1$, $\Psi=0$ or $0<\Psi<1$  if the cell is located inside the particle, in the fluid phase or is cut by the interface. Taking average of the phase indicator function in the streamwise and spanwise (homogenous) directions followed by a time averaging,  we obtain the particle concentration profile across the channel. We normalise the profile such that its mean value is equal to the total particle concentration. Although the particle centre cannot get closer than a particle radius to the wall, we still have non-zero values of the indicator function close to the wall, and thus non-zero local volume fraction.}
The wall-normal profile of local mean solid volume fraction $\phi(y)$ is displayed in figure \ref{fig:localphi}(a) for the five suspensions considered here. Note that the horizontal axis is shown in logarithmic scale to ease the comparison in the near-wall region. 
For all cases the local concentration is characterised by a homogenous particle distribution  in the bulk of the flow due to the action of turbulent mixing.
A different behavior is observed in the near-wall region where an inhomogeneous distribution occurs. 
In particular, we find a local maximum of $\phi(y)$ close to the wall, lower than the value of the bulk concentration, followed by a local minimum. In mono-disperse suspensions, the location of the maximum occurs at a wall distance slightly larger than a particle radius, while the minimum of $\phi$ occurs around one particle diameter away from the wall. 
This behaviour is attributed to the formation of a near wall layer induced by the planar wall symmetry and the excluded finite volume of the rigid particles, as noted in \cite{Picano15}. 
Bi-disperse suspensions show a progressive modification from the case of mono-disperse suspensions of small particles to that of large particles; in particular the location of the first maximum of $\phi(y)$ indicates that smaller particles show a higher accommodation at the wall. The bi-dispersity however tends to smear out the near-wall volume fraction profile.

In figure \ref{fig:localphi}(b) we show, separately, the concentration profile of small and large particles for all the cases. We note that the smaller particles tend always to form a layer, while layering is absent for the larger particles when these are in a bi-disperse suspensions and for lower values of their volume fraction. 
To better understand the effect of the bi-dispersity on the particle layering we show in \ref{fig:localphi}(c) the local volume fraction of small and large particles  normalized by their bulk values. It is clear from this figure that bi-dispersity tends to reduce or even destroy the wall-layer of the larger particles while promoting the layering of the smaller particles near the wall. Hence for bi-disperse cases the near-wall dynamics tends to be controlled by the smaller particles that concentrate more at the wall. As we will discuss in analogy with the work by \cite{Costa16}, the near-wall layering dynamics plays a crucial role in determining the overall suspension drag.

Segregation of small and large particles in different spatial locations is a well-known phenomenon and has been already explored in the seminal experimental study of \cite{Bagnold54}. Performing an experiment on the slope of falling sand with mixed sizes, it is observed that the smaller grains tend to migrate towards regions of highest shear while the larger particles accumulate in regions of lower shear. 
This has been attributed to the proportionality of the disperse phase pressure to the square of the grain diameter. Larger particles will be pushed away from regions of high shear rate  by the stronger dispersive pressure {\cite[see][for recent theoretical analysis on this aspect]{Schlick16}}. Even though a turbulent flow tends to redistribute the particles and to weaken this effect, we still observe segregation of small/large particles at the wall/centre region of the channel, an observation consistent with the findings by Bagnold. To further support this, 
we discuss the concentration profile of binary mixture suspensions in laminar flows, $Re_b=1000$, in Appendix A.

\begin{figure}
\begin{center}
   \includegraphics[width=7.0cm]{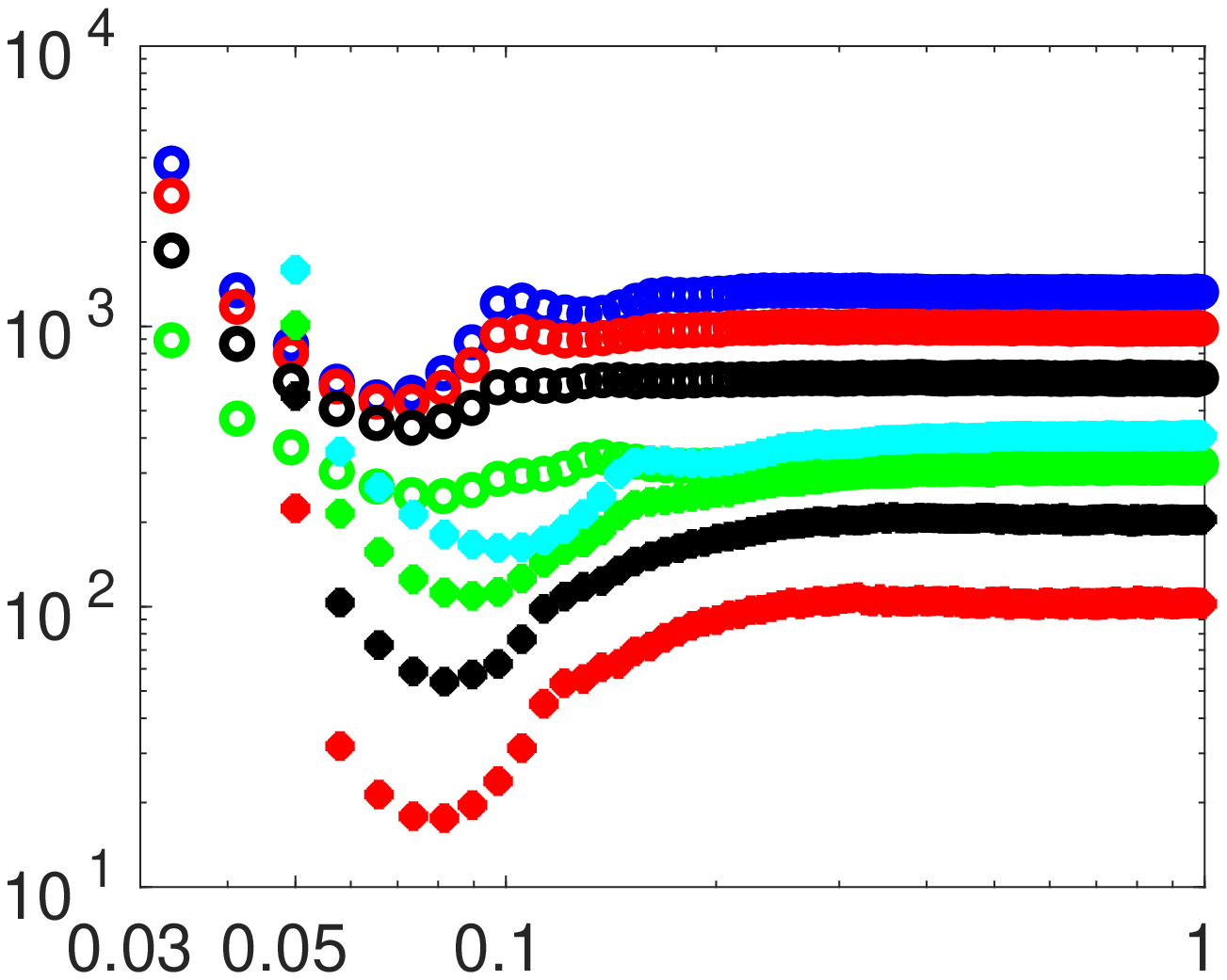}
  \put(-185,155){{\large $(a)$}} 
  \put(-215,90){{\large $\bar n$}} 
  \put(-100,-5){{\large $y/h$}} 
   \includegraphics[width=7.0cm]{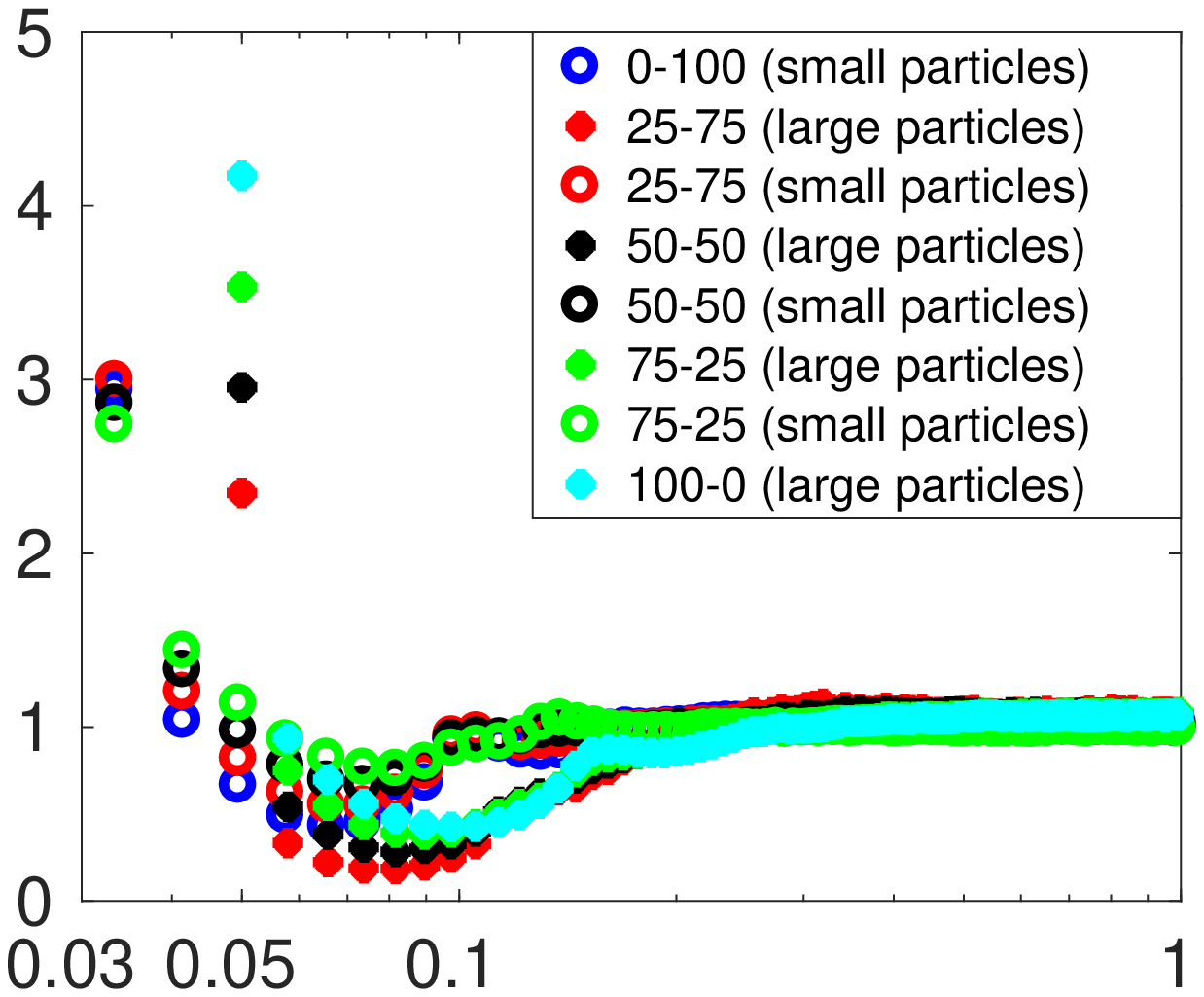}
  \put(-185,155){{\large $(b)$}} 
   \put(-15,90){{\large $\bar n/n_0$}} 
    \put(-100,-5){{\large $y/h$}} 
   \caption{a) Profiles of the mean particle number density, $\bar n$, b) normalized mean particle number density, $\bar n/n_0$, across the channel at $Re_b=5600$}
   \label{fig:n_density}
\end{center}
\end{figure}

{Here, we also examine the particle number density across the channel as an alternative measure of the particle distribution. In this case, we consider only particle centres and compute the spatial and time average of their wall normal positions. In figure \ref{fig:n_density}(a), we show the profile of the mean particle number density, $\bar n$, of small and large particles for the five turbulent cases studied. This is computed by counting the mean number of particle centres at each wall normal station per unit volume. For each case, the peak of the mean number density occurs at a distance to the wall equal to the particle radius. Small particles are predominant close to the wall in cases $25-75$ and $50-50$. This is not true for the case $75-25$ where the number of small and large particles close to the wall is of the same order. This result is inline with that of the local volume fraction reported in the previous figure. In panel (b) we normalise the data of panel (a) by the corresponding bulk value, $\bar n / n_0$. Similarly the high traffic of small particles close to the wall is evident.}

\subsection{Overall drag}
The mean velocity profile and the overall drag  of a turbulent channel flow are directly connected to the inhomogeneous multi-scale dynamics of the turbulent flow. In wall turbulence, it is common to define a bulk Reynolds number, $Re_b$, representing a dimensionless form of the flow rate and the friction Reynolds number, $Re_\tau=u_\tau h /\nu$, as a dimensionless measure of the overall drag. Based on well established empirical correlations, \cite[e.g.][]{Pope00}, it is possible to relate the two numbers for the case of the unladen turbulent plane channel flow,
\begin{equation}
Re_\tau= u_{\tau}h/\nu = 0.09(Re_b)^{0.88}.  
\label{eq:retreb}
\end{equation}
When a turbulent flow with suspended particles is considered, the relation (\ref{eq:retreb}) needs to be modified to consider the rheological properties of
the suspension. Laminar suspensions are characterised by a monotonic increase of their effective viscosity as a function of the volume fraction of the dispersed phase. The effective viscosity of the suspension in laminar condition has been first predicted in the seminal work by \cite{Einstein06} for the dilute regime in the form of a linear correction, $\nu_e= \nu (1+2.5\Phi)$. Later on \cite{Batchelor72} considered pair interactions and derived a second-order correction in the volume fraction valid for semi-dilute suspensions. In the dense regime, however, only empirical fits are available, among others the Eilers fit,
\begin{equation}
\nu_e= \nu (1+ 5/4 \frac{\Phi}{1-\Phi/\Phi_{max}})^2, 
\label{equ:EF}
\end{equation}
 where $\Phi_{max} \approx 0.6$ is the maximum packing fraction of particles \cite[][]{Stickel05}. {
 Note that \cite{Mwasame16} have recently  introduced a weighting function $\beta$, describing the effects of size ratio and volume-fraction ratio on the effective viscosity of binary suspensions. Interestingly, this model shows that for concentrations $\Phi<0.3$, the effect of particle bi-dispersity on the effective viscosity of the suspension is negligible, i.e.\ $\beta \approx 1$. Thus, in our study at $\Phi=0.2$, we safely estimate the effective viscosity using equation \ref{equ:EF}.}
When inertia dominates the dynamics, deviations from the empirical fits valid in viscous regimes have been reported by several authors, e.g. inertial shear-thickening \cite[e.g.][]{Morrispof08,Maxey11,Picano13}. 
 
\begin{figure}
\begin{center}
   \includegraphics[width=9cm]{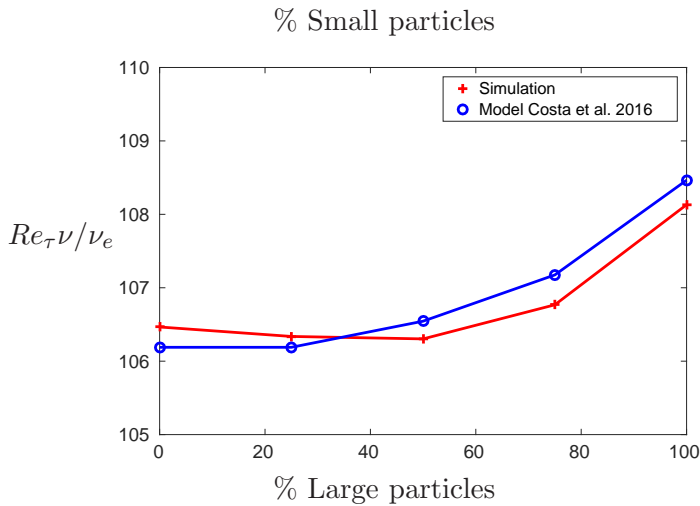}
   \put(-280,90){{\large $Re_\tau \nu /\nu_e$}}
    \put(-180,-5){{\large $\%$ Large particles}}
    \put(-180,170){{\large $\%$ Small particles}}
   \caption{Effective friction Reynolds number pertaining the different cases $Re^{e}_\tau$ considered here and the predictions by the model of \cite{Costa16}.} 
   \label{fig:Ret}
\end{center}
\end{figure}

The situation becomes even more complicated in turbulent flows laden with particles larger than the smallest hydrodynamic scales. As discussed in \cite{Picano15,Prosperetti15}, the turbulent friction Reynolds number cannot be predicted by only taking into account the effective viscosity of the suspension, 
\begin{equation}
Re^{e,exp}_\tau=0.09(Re_b \nu/\nu_e)^{0.88},
\label{eq:exp_re}
\end{equation}
 where $Re^{e,exp}_\tau$ is the expected suspension friction Reynolds number which accounts for the suspension dynamics only by considering the suspension effective viscosity.  
Applying the formula above, the friction Reynolds number pertaining the cases considered here would be $Re^{e,exp}=101.7$, obtained using $\nu_e\simeq1.9\nu$. This should be
compared with the values of the  effective friction Reynolds number $Re^{e}_\tau=u_{\tau}h/\nu_e$ extracted from DNS of the present bi-disperse cases and reported in figure \ref{fig:Ret} where all  values are higher than what predicted from eq.~\eqref{eq:exp_re}. 
In addition, the flow shows a dependence on the particle size which is not accounted for by eq.~\eqref{eq:exp_re}. As shown here and in the previous works mentioned above, the presence of finite-size spherical particles increases the overall drag more than what predicted considering only the effective viscosity. An interesting exception is provided by the case of oblate disc-like particles, see \cite{Niazi16}.

To explain these observations, a new theoretical framework has been recently  proposed by \cite{Costa16} to extend the law of the wall for the turbulent channel flow of finite size mono-disperse rigid spheres. The underlying idea is that the formation of a particle layer at the wall creates an additional source of drag.
In the formula proposed  by these authors, the overall stress is assumed to depend on the Reynolds number, the volume fraction and the particle size that together determine the wall layer thickness \cite[for the details of the derivations we refer the readers to][]{Costa16}. 
The formula reads   
\begin{align}
Re^{S}_\tau=Re^{e,exp}_\tau \bigg{(}1-\frac{\delta_{pw}}{h}\bigg{)}^{-3/2+0.88},
\label{eq:ReS}  
\end{align}
where the particle layer thickness $\delta_{pw}=C(\Phi/\Phi_{max})^{(1/3)}D_p$ and $C \approx 1.5$. This has been found to provide a good fit of data form simulations of particle-laden turbulent channel flow for a reasonable range of volume fractions and Reynolds numbers.

{To extend this formula to bi-disperse suspensions we need to provide an ''effective" particle size in order to characterize the wall layer.  Examining the local volume fraction profiles in figure~\ref{fig:localphi}, we note that the distance between the local maximum and the wall is proportional to the particle diameter for mono-disperse cases. In our simulations the constant of proportionality is obtained equal to $1.37$. We therefore assume the effective particle size, $D_p^{e}$, of the binary mixtures as the product of the location of the local maximum in the concentration profile with the constant of proportionality, $1.37$. In table \ref{tab:tab2} we report the values of the effective particle diameter pertaining each simulation normalised with half the channel height.}

\begin{table*}
  \begin{center}
    \setlength\tabcolsep{4ex}
  \begin{tabular*}{\textwidth}{*{6}{c}}
    \hline
    Case name               & $0-100$ & $25-75$ & $50-50$ & $75-25$ & $100-0$ \\ \hline
    $D_p^{e}/h $     &   0.066    &  0.066  &   0.071    &   0.081  &   0.1       \\ \hline  
  \end{tabular*}
  \caption{Effective particle diameter based on the location of local maximum in the profile of the local concentration}
  \label{tab:tab2}
\end{center}
\end{table*}

{Using this definition, we note that  $D_p^{e}$  almost coincides with the smallest particle diameter for the cases $0-100$, $25-75$ and $50-50$; it then  increases and reaches the largest particle diameter for the $100-0$ case, as expected. In figure~\ref{fig:Ret}, we report the effective friction Reynolds number $Re^{S}_\tau$ estimated from eq.~\eqref{eq:ReS} with $C\simeq 1.33$. We note that the predicted values are in very good agreement with the DNS data. This analysis also explains the trend for the variation of the overall drag observed in bi-disperse suspensions when increasing the percentage of large particles.}

 \subsection{Turbulent statistics}

\begin{figure}
\begin{center} 
   \includegraphics[width=.32\textwidth]{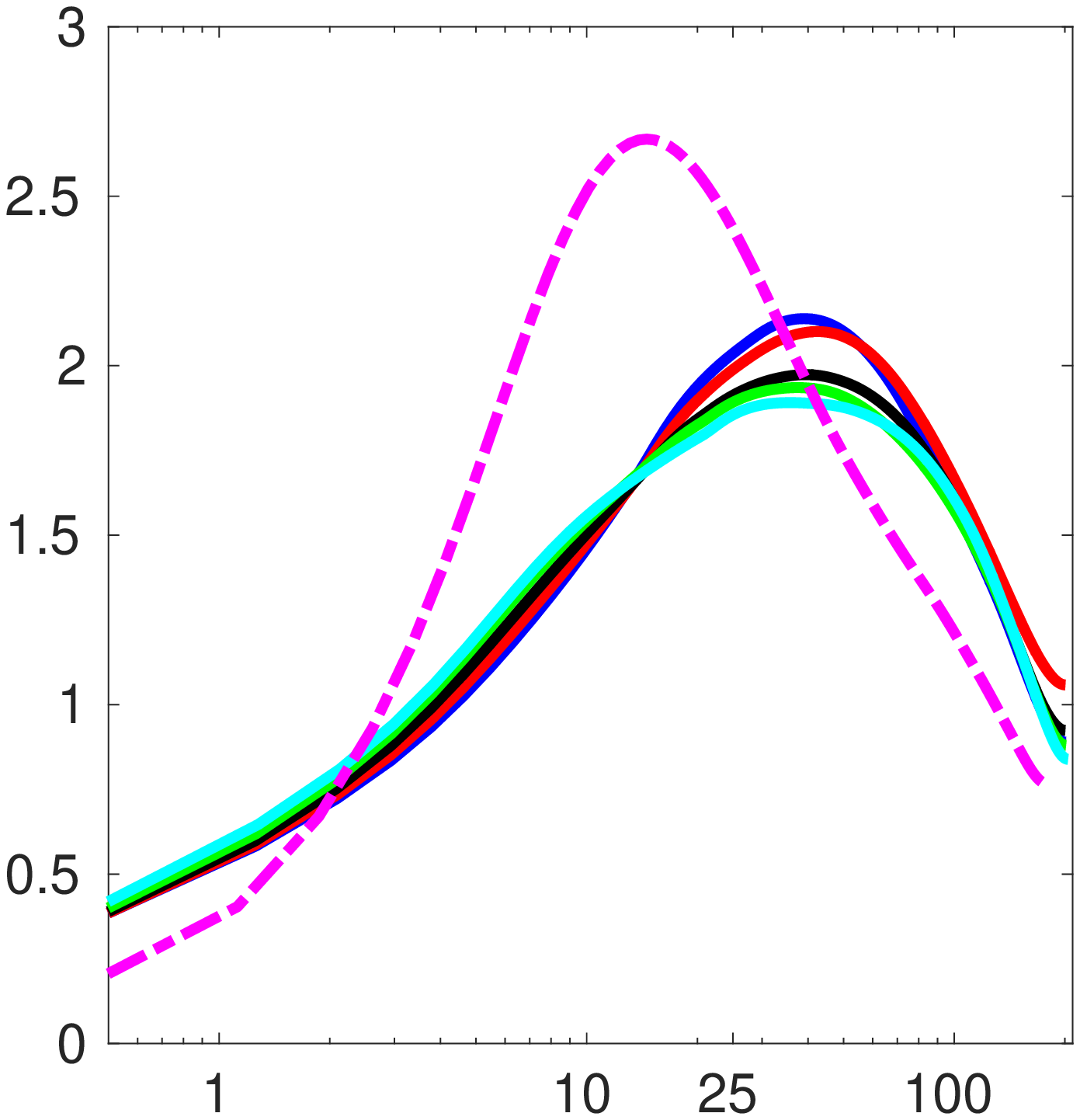}
   \put(-125,120){{$(a)$}} 
    \put(-100,85){{  $u'^+_{f}$}} 
    \put(-70,-2){{  $y^+$}} 
   \includegraphics[width=.32\textwidth]{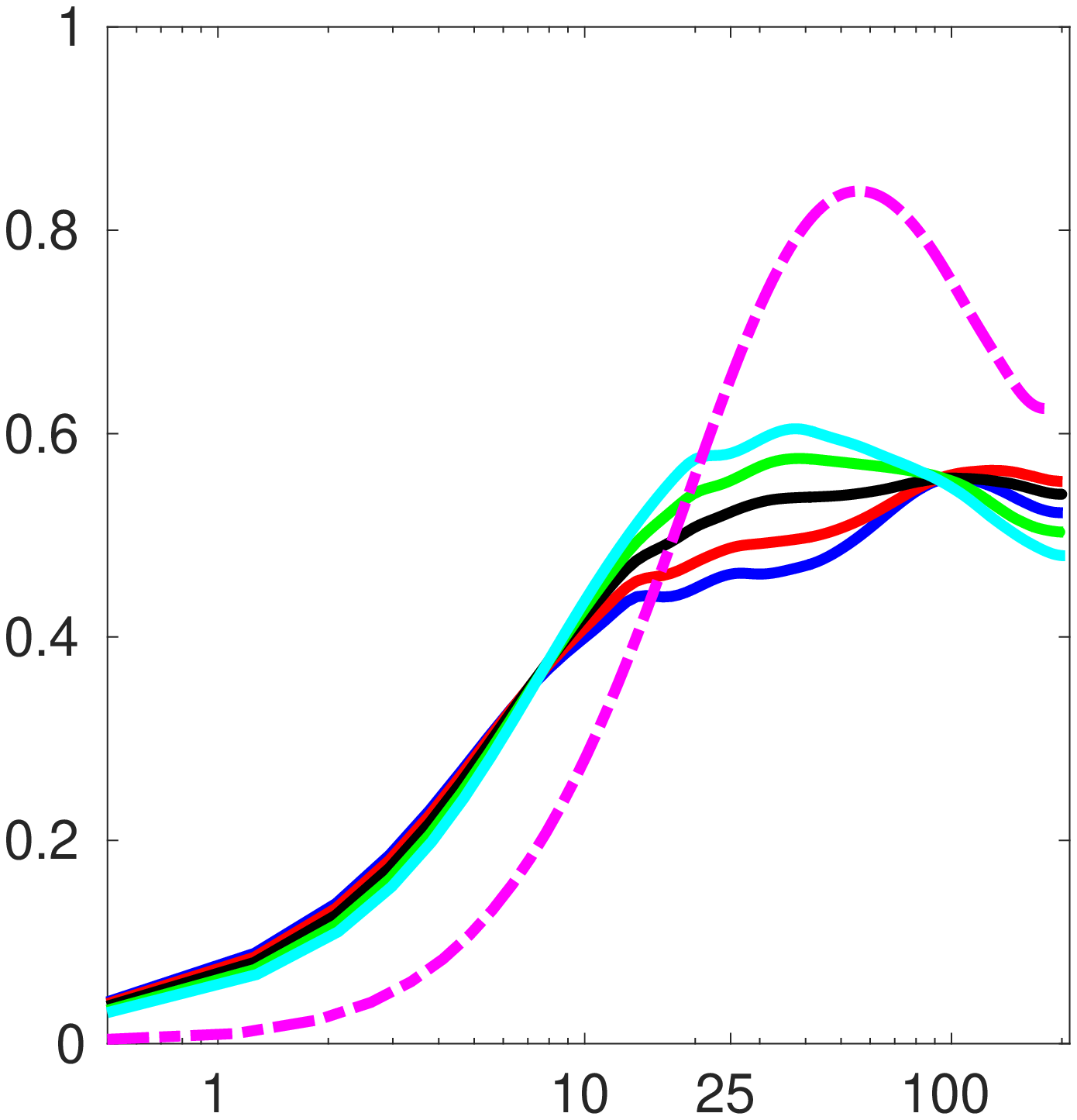}
    \put(-125,120){{  $(b)$}} 
     \put(-100,85){{  $v'^+_{f}$}} 
     \put(-70,-2){{  $y^+$}}
   \includegraphics[width=.32\textwidth]{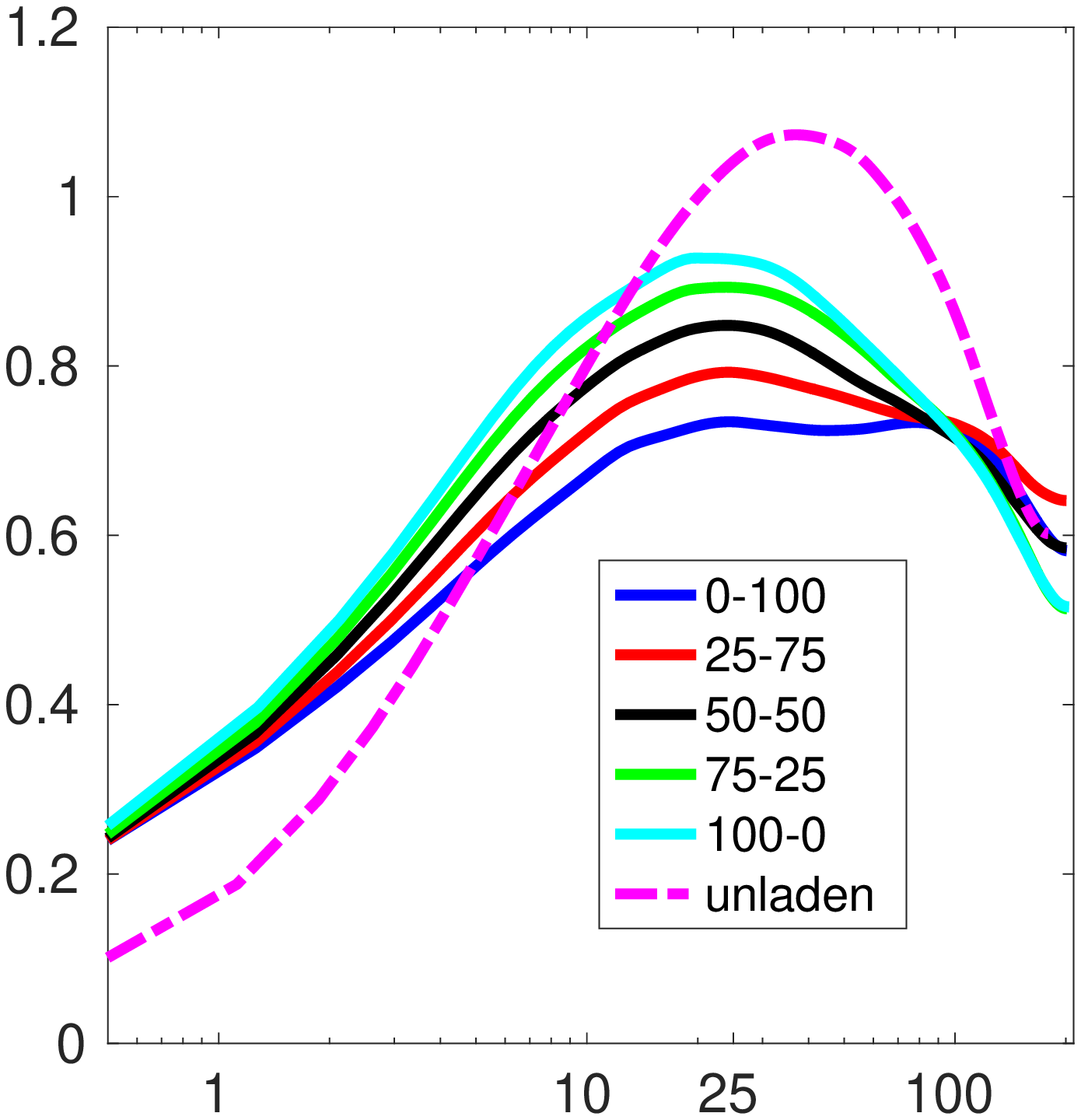}
    \put(-125,120){{  $(c)$}} 
     \put(-100,85){{  $w'^+_{f}$}} 
     \put(-70,-2){{  $y^+$}} 
     \caption{Profiles of the intensity of the fluid velocity fluctuation components in (a) streamwise (b) wall-normal and (c) spanwise directions scaled in inner units. The data pertain the different cases under investigation as indicated in the legend.}
   \label{fig:rmsplus}
\end{center}
\end{figure}

Next we examine the statistics of the turbulent flow. Figure \ref{fig:rmsplus} shows the root mean square (r.m.s.) of the fluid velocity fluctuations for the five particulate cases in inner units, together with the corresponding statistics for the unladen turbulent channel flow at the same bulk Reynolds number, $Re_b=5600$. Comparing the statistics of the particulate and the unladen flow we observe that the turbulence activity reduces in all the particulate flows, with lowered peaks. The mechanisms responsible for this have been discussed in detail in the previous studies by \cite{Lashgari14} and \cite{Picano15} for mono-disperse suspensions: this reduction is connected to the increasing importance of particle stresses with respect to Reynolds stresses in  the momentum transfer. The profiles of the streamwise velocity fluctuations $u'^+$, displayed in \ref{fig:rmsplus}(a), reveal that the wall-normal location of the peak is shifted toward the channel centreline for the particulate flows. This can be explained by the formation of the particle layers close to the wall that hinders the production of  perturbation kinetic energy from the mean flow. The maximum streamwise velocity fluctuation is reduced for all cases and appear to be slightly higher for bi-disperse suspensions with larger percentages of small particles. The opposite behaviour is observed for the cross stream velocity fluctuations, \ref{fig:rmsplus}(b,c) where the peaks move toward the wall with respect to the unladen flow. For these components, we note the opposite trend  as that of the streamwise fluctuations when varying the relative amount of large to small particles. We infer that large particles arriving to or departing from the wall layer induce higher level of fluctuations in the cross-stream directions.  Finally, we note that very close to the wall higher fluctuations are present in the particle-laden cases, which is attributed to the relatively large particle slip velocity and the squeezing motion of the fluid between the wall and the particles.

\begin{figure}
\begin{center}
   \includegraphics[width=.32\textwidth]{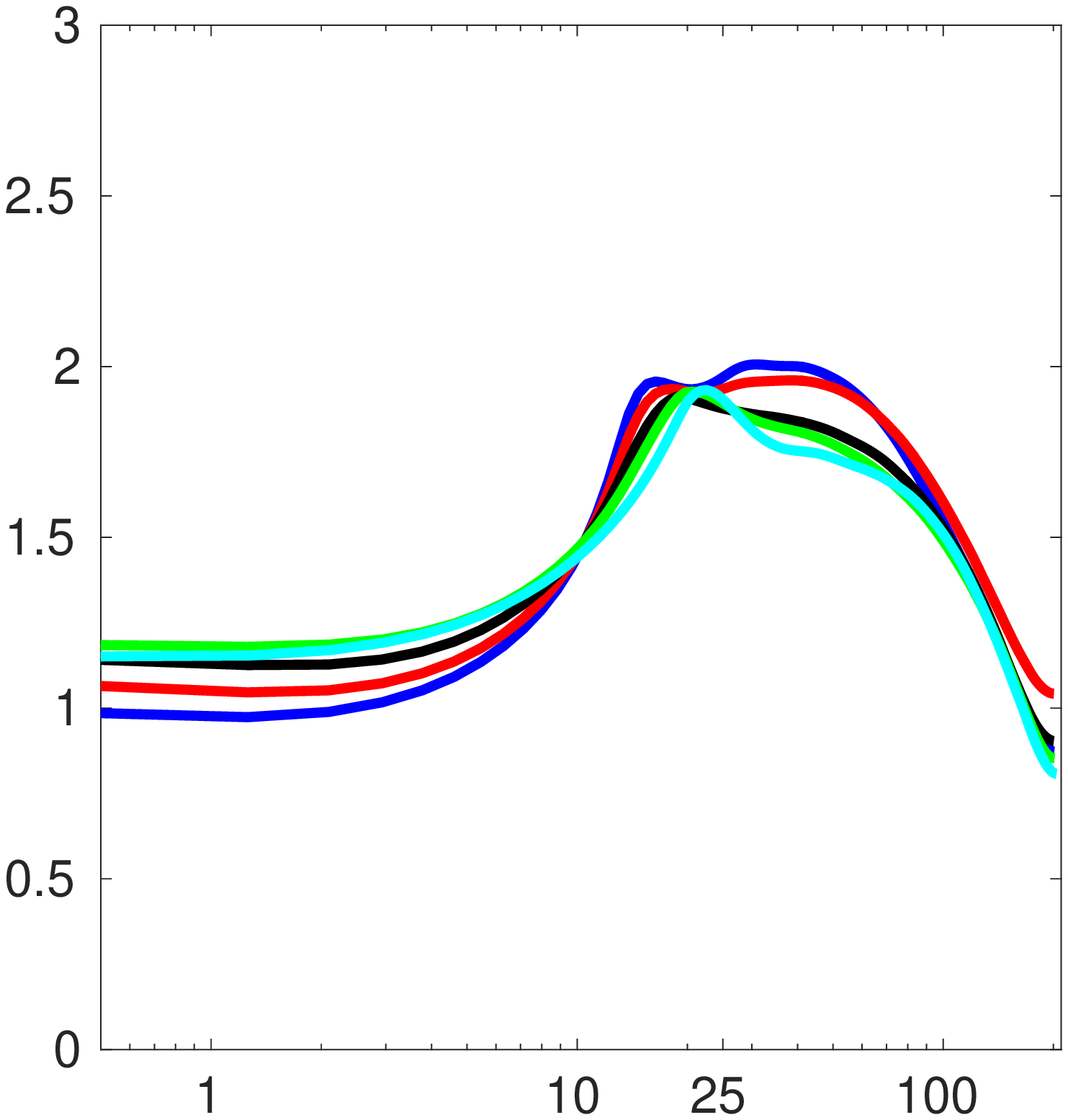}
   \put(-125,120){{$(a)$}}
    \put(-100,85){{  $u'^+_{p}$}} 
    \put(-70,-2){{  $y^+$}} 
   \includegraphics[width=.32\textwidth]{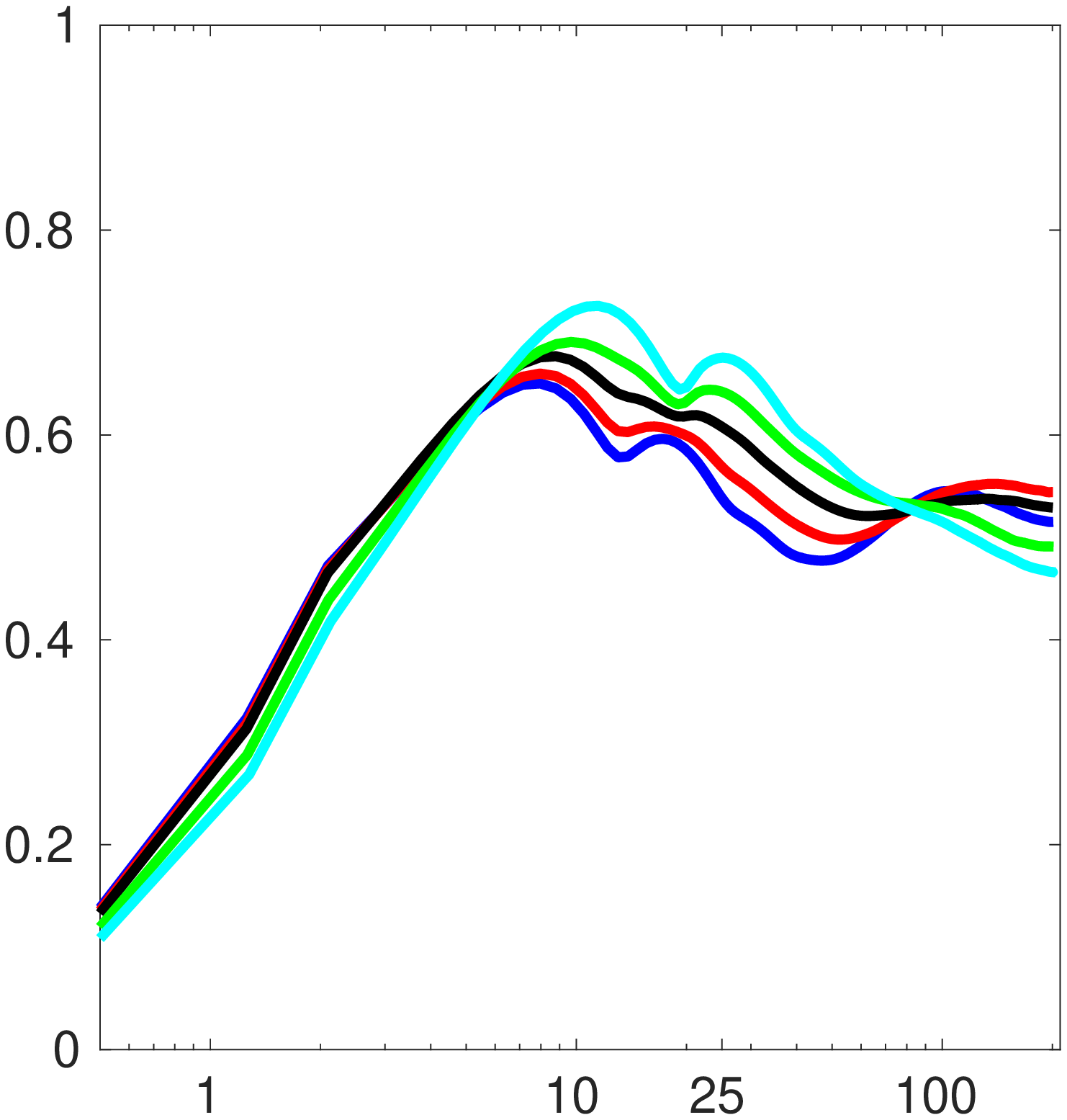}
     \put(-125,120){{  $(b)$}} 
     \put(-100,85){{  $v'^+_{p}$}} 
     \put(-70,-2){{  $y^+$}}
    \includegraphics[width=.32\textwidth]{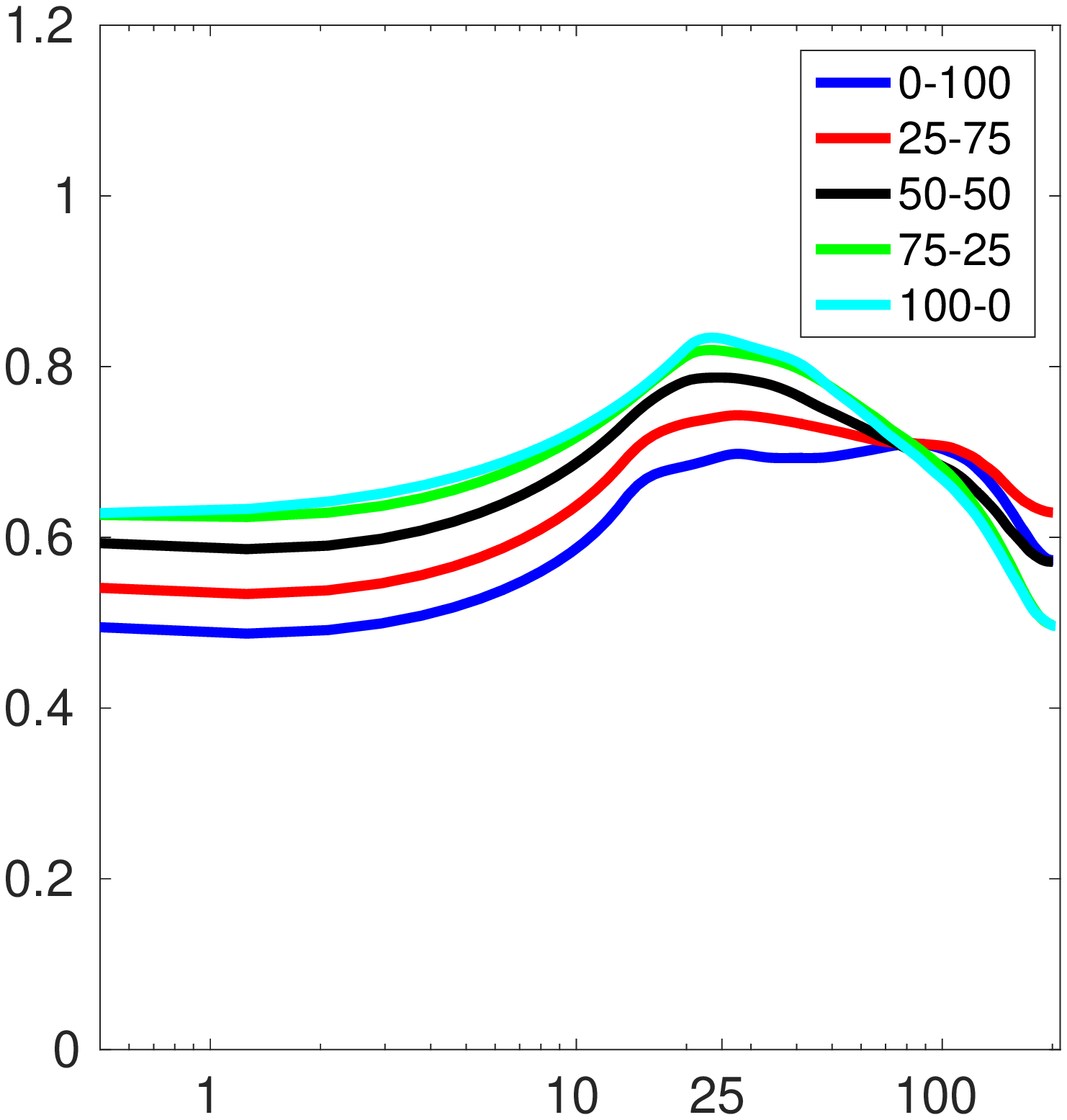}
     \put(-125,120){{  $(c)$}} 
     \put(-100,85){{  $w'^+_{p}$}} 
     \put(-70,-2){{  $y^+$}} 
     \caption{Profiles of the intensity of the particle velocity fluctuation components in (a) streamwise (b) wall-normal and (c) spanwise directions scaled in inner units. The data pertain the different cases under investigation as indicated in the legend.}
   \label{fig:prmsplus}
\end{center}
\end{figure}

The r.m.s. velocity fluctuations of the particle phase are shown in figure \ref{fig:prmsplus} in inner units in order to have a direct comparison with those pertaining the  fluid phase. First, we observe that the fluctuations do not vanish at the wall,  unlike those of the fluid phase with the only exception of the wall-normal component. In the region close to the wall where the first layer of particles is formed the streamwise and cross-stream fluctuations of the case with all large/small particles are the largest/smallest, see figure \ref{fig:prmsplus}(a,b,c). This is attributed to the higher/lower slip velocity of the large/small particles in the near-wall region. The opposite behaviour is observed in the core of the channel where small particles are subjected to more agitation due to the turbulent activity. 
Away from the near-wall region, the level of fluctuations of the particle phase is generally lower than that of the fluid phase.

 \subsection{Particle dynamics}
 
In this last part we focus on the particle dispersion dynamics by measuring the particle lateral displacement, which results from particle-particle and particle-fluid interactions. In this work we examine only the particle mean square displacement in the spanwise direction to avoid inhomogeneous and mean flow effects that determine the motion in the wall-normal and streamwise directions. We analyse separately the behavior of small and large particles in the binary suspensions. The mean square displacement in the spanwise direction is defined by $\langle \Delta {z}^2_p  \rangle (\Delta t)= \langle [ \textbf{z}_p(t+ \Delta t) - \textbf{z}_p(t) ]^2 \rangle_{p,t} $ where $z_p$ is the vector containing the spanwise position of the particle centres and $\Delta t$ is the time interval. The ensemble average, $\langle  \rangle_{p,t}$ is taken over all the particles and times after a fully-developed flow is established. For more details about particle dispersion and diffusion we refer the readers to the work by \cite{Hinch96} and \cite{Sierou04}.

\begin{figure}
\begin{center}
   \includegraphics[width=.5\textwidth]{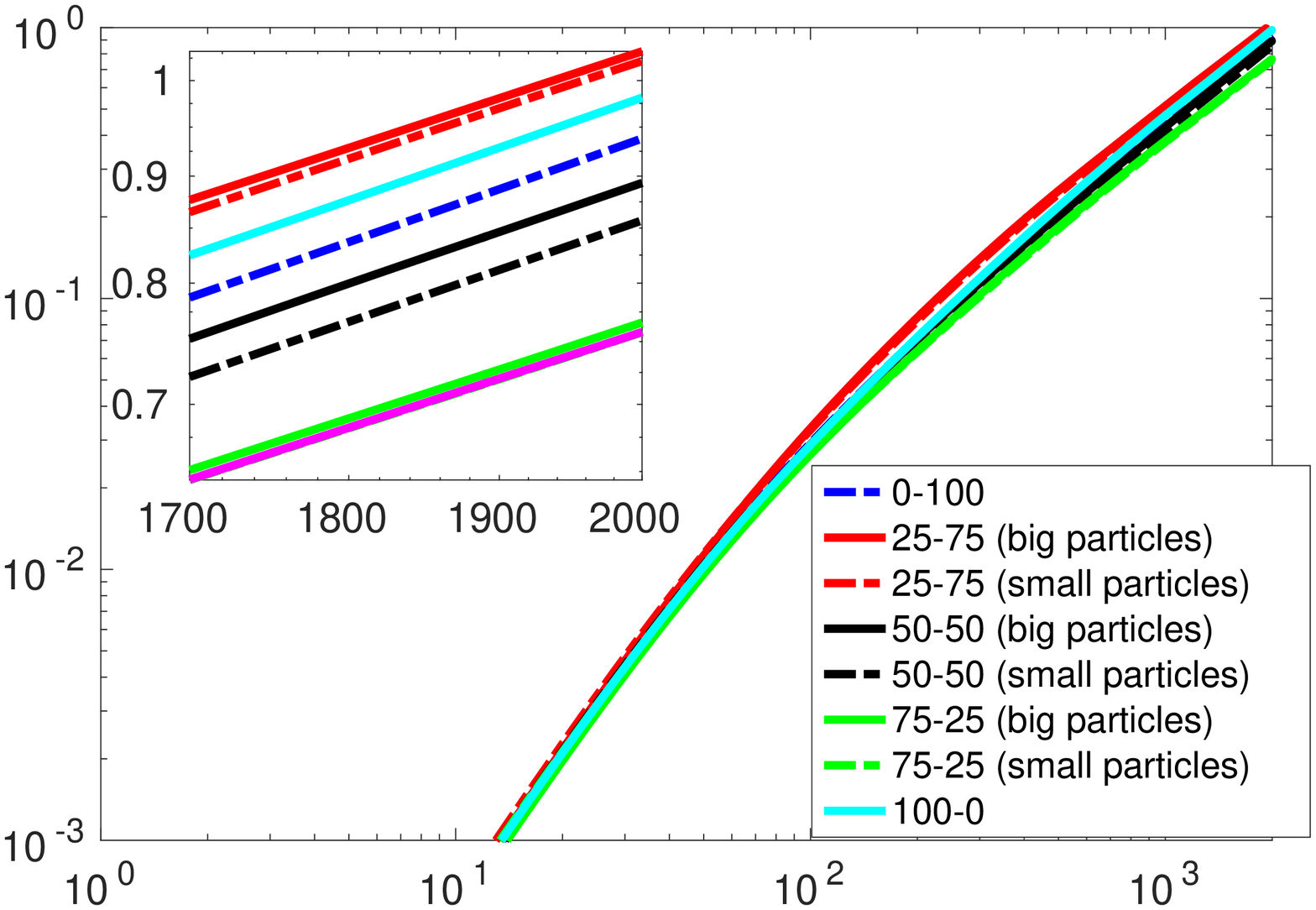}
    \put(-190,120){{\large$(a)$}}
    \put(-190,60){{\begin{rotate}{90} $\langle  \Delta z^2_p \rangle$ \end{rotate}}} 
     \put(-110,-5){{ $\Delta t$}} 
     \includegraphics[width=.5\textwidth]{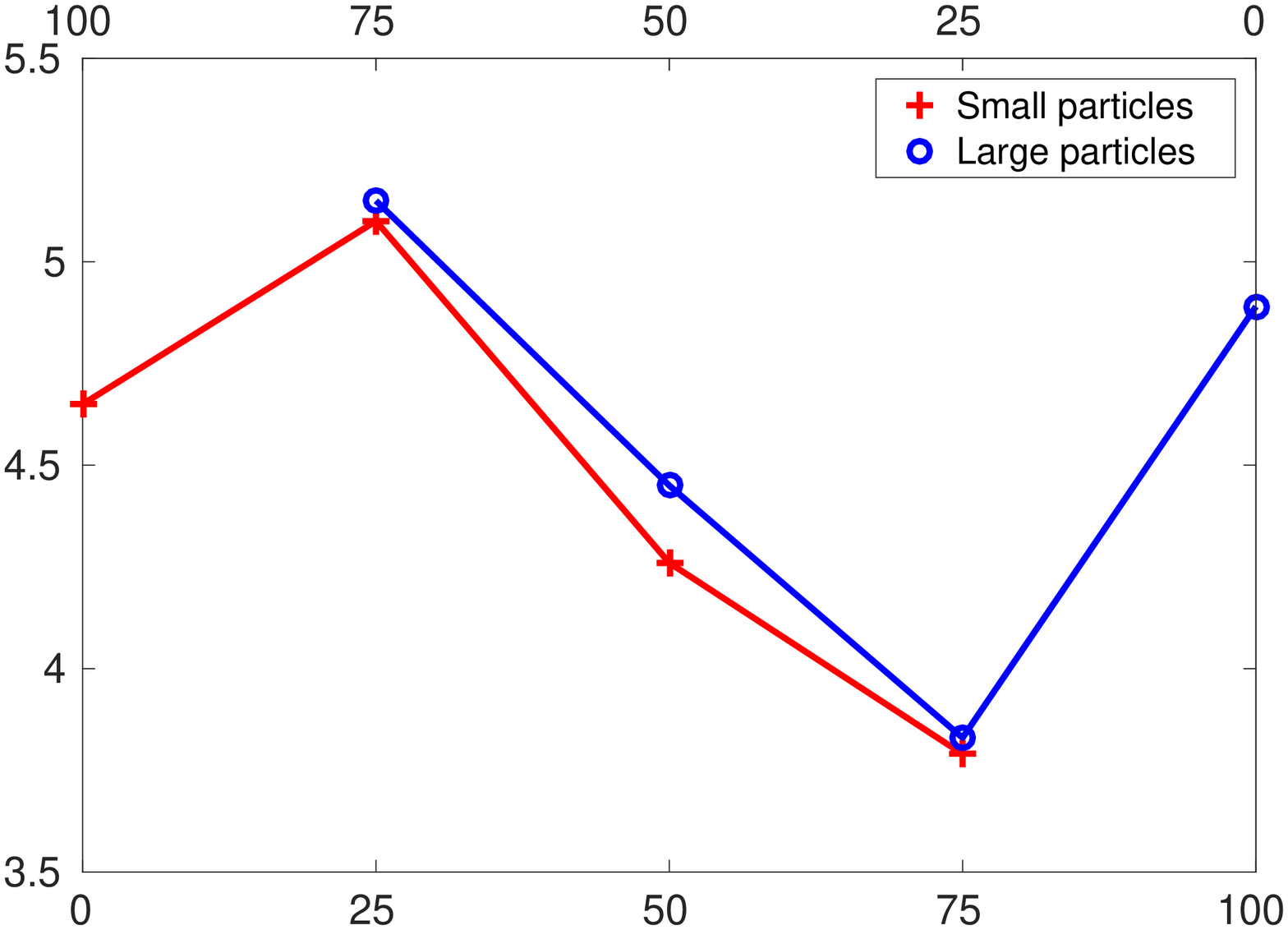} 
     \put(-190,120){{\large$(b)$}}   
    \put(-190,60){{ \begin{rotate}{90} $D_{zz}.10^4$ \end{rotate} }}
    \put(-140,-5){{$\%$ Large particles}}
    \put(-140,130){{$\%$ Small particles}}
   \caption{a) Particle mean square displacement $\langle \Delta z^2_p \rangle$ for the different cases under investigation. b) Dispersion coefficient, $D_{zz}$, in the spanwise direction versus the relative particle volume fraction. }
   \label{fig:dis}
\end{center}
\end{figure}

The particle mean square displacement is shown in figure \ref{fig:dis}(a) versus time. {Note that the mean square displacement,  $\langle \Delta z^2_p \rangle$, is normalized by $(2h)^2$ whereas the time is expressed in units of $t_{ref}=d_l/U_b$}. For all the cases,  as expected, the particle mean square displacement varies  initially quadratically in time,  $\langle \Delta {z}^2_p (\Delta t) \rangle \propto \Delta t^2 $, indicating a high correlation in the particle trajectories at small intervals. Later on, the classical diffusive behaviour takes over: the particle trajectories de-correlate and the mean square displacement varies linearly with time. In the inset of the figure we report the regime behavior, the linear diffusive part, for the different particle-laden flows. 

To better highlight the differences we provide in the panel b) of figure \ref{fig:dis}, the value of the dispersion coefficient, $D_{zz}$, obtained by fitting the diffusive, long time regime to a straight line. As appreciated from the figure, a non-monotonic behaviour is observed. Monodisperse suspensions of either small or large particles show a similar value of the diffusion coefficient. In bidisperse suspensions, interestingly, the diffusion coefficient of large and small particles is always close, with a slightly higher value for the largest particles. 
We explain this non-monotonic behavior by the following arguments. 
Adding a small amount of bigger particles in a monodisperse suspension of small particles, as in the 25-75 case, we expect that the disturbances induced by the large particles enhance the particle dispersion. Conversely, adding a small amount of smaller particles to a monodisperse suspension of large particles, as in the 75-25 case, will reduce the overall dispersion because of the reduced mobility induced by the smaller particles that are positioned among the larger ones. The peculiar picture of the diffusion coefficient for bi-disperse suspension can be explained considering that these opposite trends need to match when the concentration of the two particles are similar. 
{Finally, we note that the non-monotonic behaviour of the dispersion coefficient is observed for the particle size distribution studied here and may need further investigations for different size distributions.}

 \begin{figure}
\begin{center}
   \includegraphics[width=.5\textwidth]{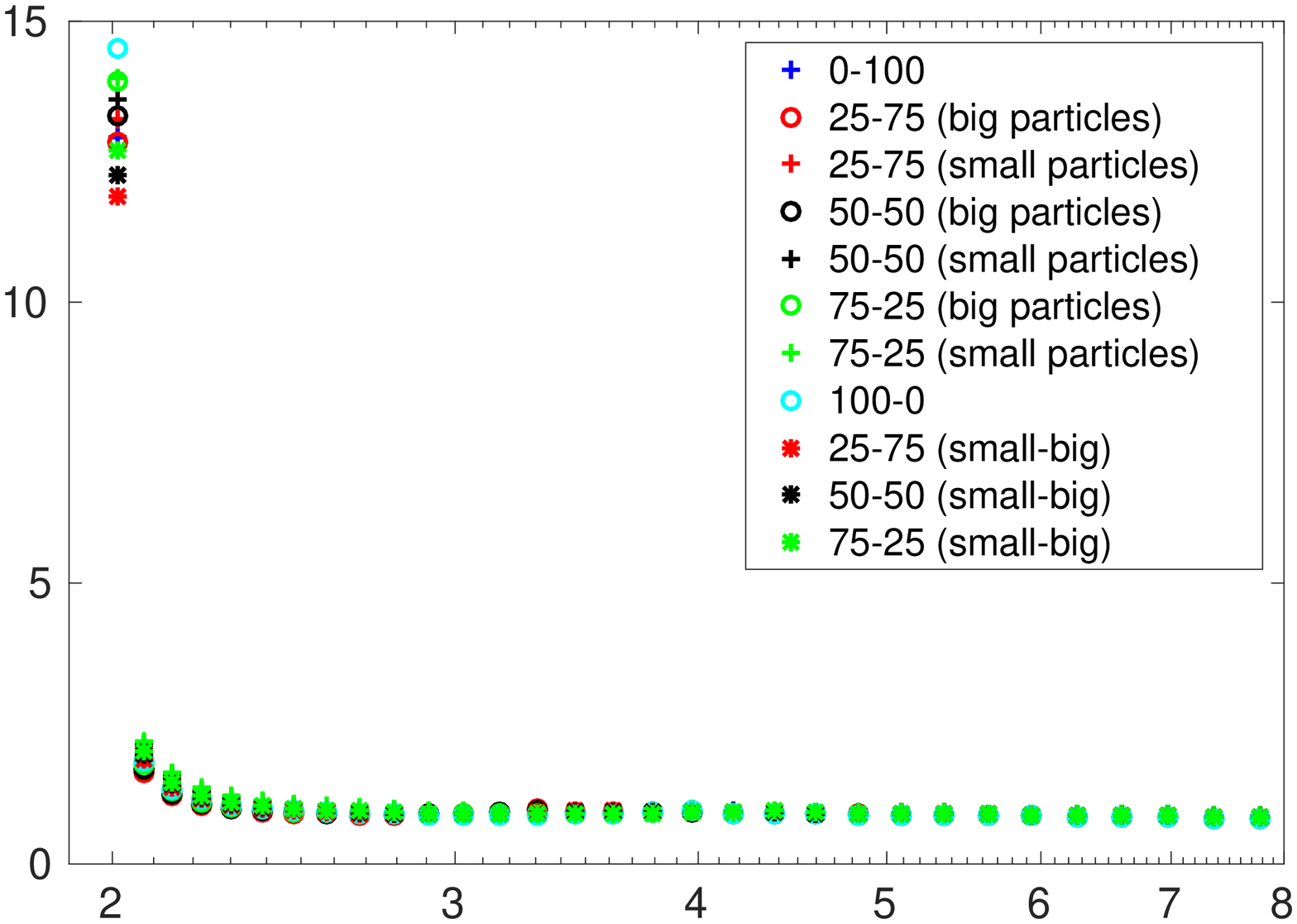}
      \put(-190,120){{\large$(a)$}}
     \put(-190,70){{\begin{rotate}{90} $g(r)$ \end{rotate}}}
      \put(-95,0){{$r/a$}}
   \includegraphics[width=.5\textwidth]{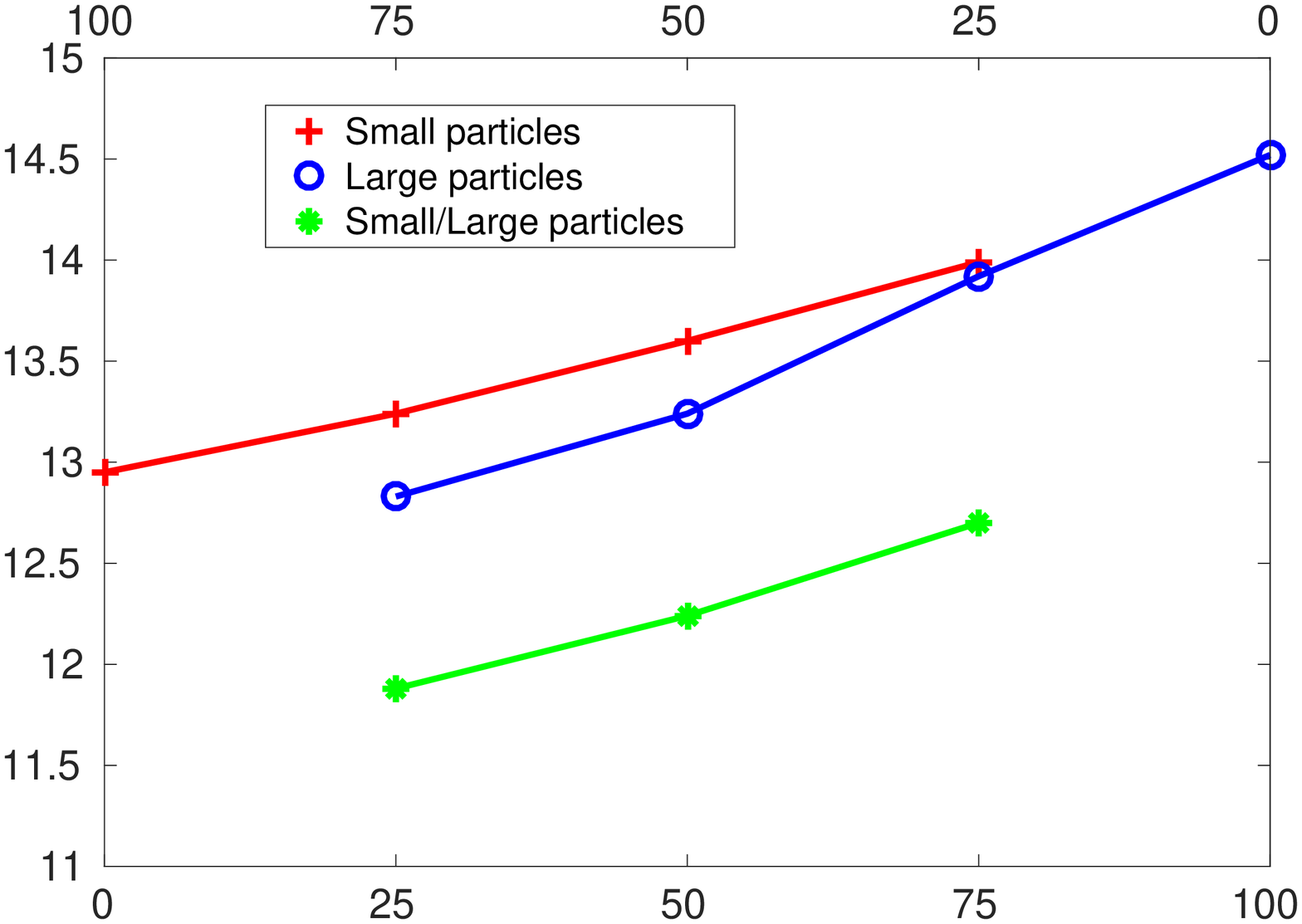}
      \put(-190,120){{\large $(b)$}} 
    \put(-190,60){{ \begin{rotate}{90} $g(r=2a)$ \end{rotate} }}
    \put(-140,-5){{$\%$ Large particles}}
    \put(-140,130){{$\%$ Small particles}}
   \caption{(a) Radial Distribution Function $g(r)$ and (b) RDF at pair contact for the five cases under consideration.}
   \label{fig:gr}
\end{center}
\end{figure}


One interesting aspect of dense bi-disperse suspensions is the probability of collision, which, in turns, depends on the particle-pair relative distribution and the first-order velocity structure functions. For binary mixtures, we report the particle-pair statistics by considering pairs consisting of i) only two small particles, ii) only two large particles and iii) one small and one large particle. The first step is to examine the Radial Distribution Function, $g(r)$, as a function of the distance between the centres of the particle pairs, $r$. 
The radial distribution function is a measure of the non-uniformity of the particle distribution and is obtained by counting the average number of particle pairs whose centres are at distance $r$ with respect to the value of a corresponding random distribution \cite[see for more details][]{Collins00,Gualtieri12}. 
In mathematical form, $g(r)=\frac{1}{A n_0} \frac{dN_r}{dr} $, where $A=4 \pi r^2$ is the area of the shell around the reference particle, $N_r$ is the number of the particle pairs within a sphere of radius $r$ and $n_0=\frac{N_p(N_p-1)}{2V}$ the density of the particle pairs in the volume $V$ with  $N_p$ the total number of particles. 

  \begin{figure}
\begin{center}
   \ \includegraphics[width=.5\textwidth]{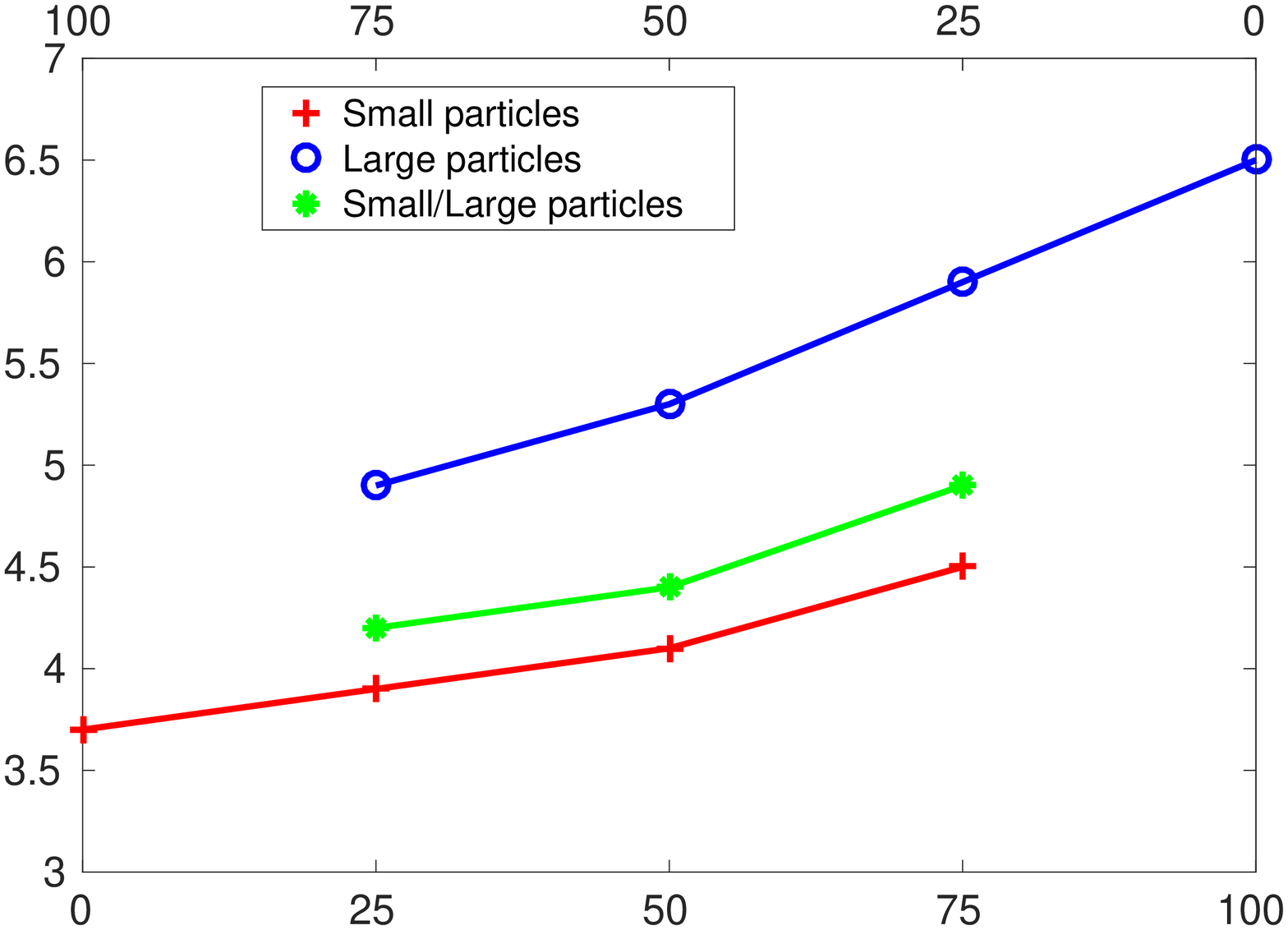}
      \put(-190,120){{\large$(a)$}}
    \put(-190,50){{ \begin{rotate}{90} $\kappa(r=2a).10^{3}$ \end{rotate} }}
    \put(-150,-5){{$\% \quad Large \quad particles$}}
    \put(-150,130){{$\% \quad Small \quad particles$}}\\
      \includegraphics[width=.5\textwidth]{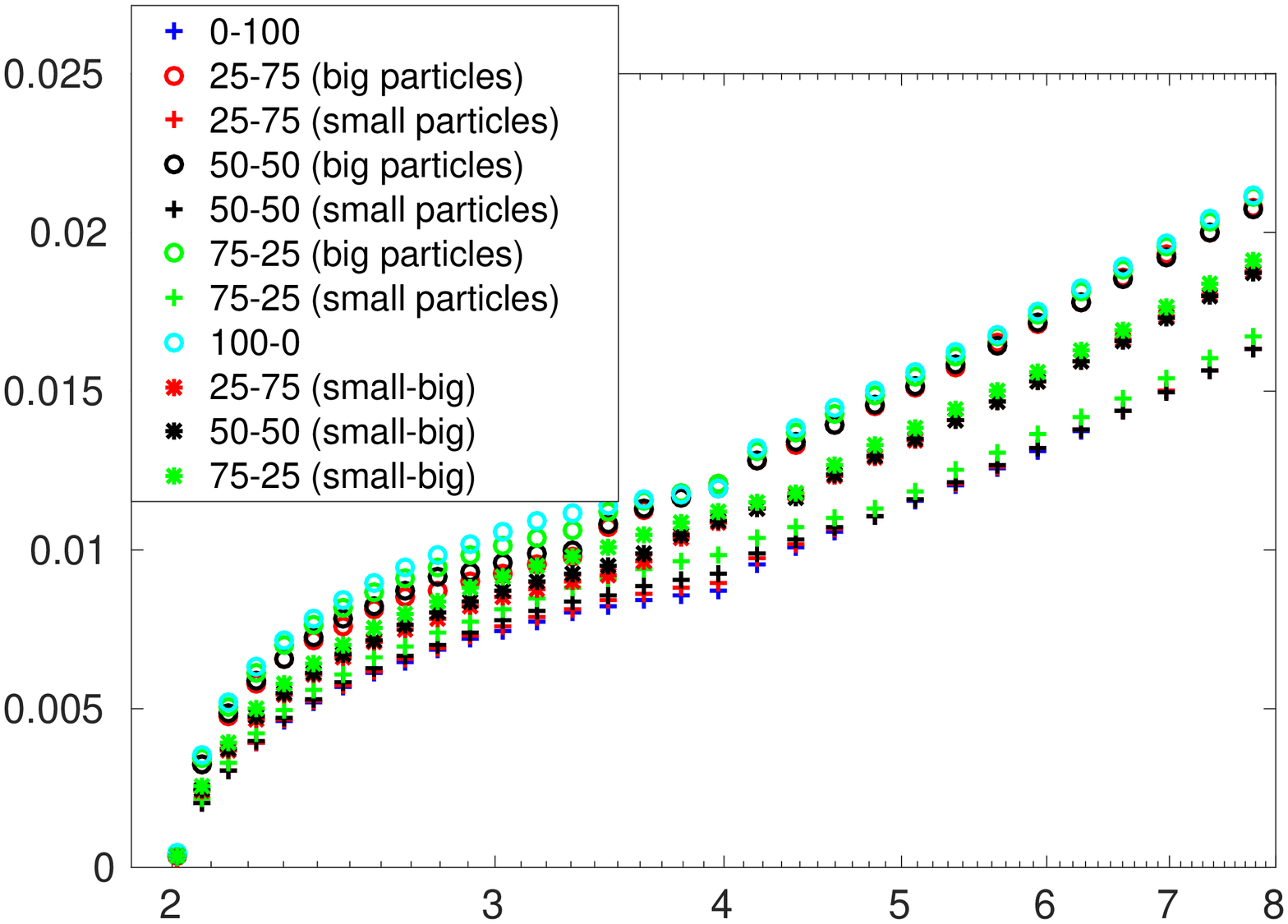}
      \put(-190,125){{\large $(b)$}} 
      \put(-190,70){{\begin{rotate}{90} $ \langle dv^-_n  \rangle$ \end{rotate}}}
      \put(-95,0){{ $r/a$}}
       \includegraphics[width=.5\textwidth]{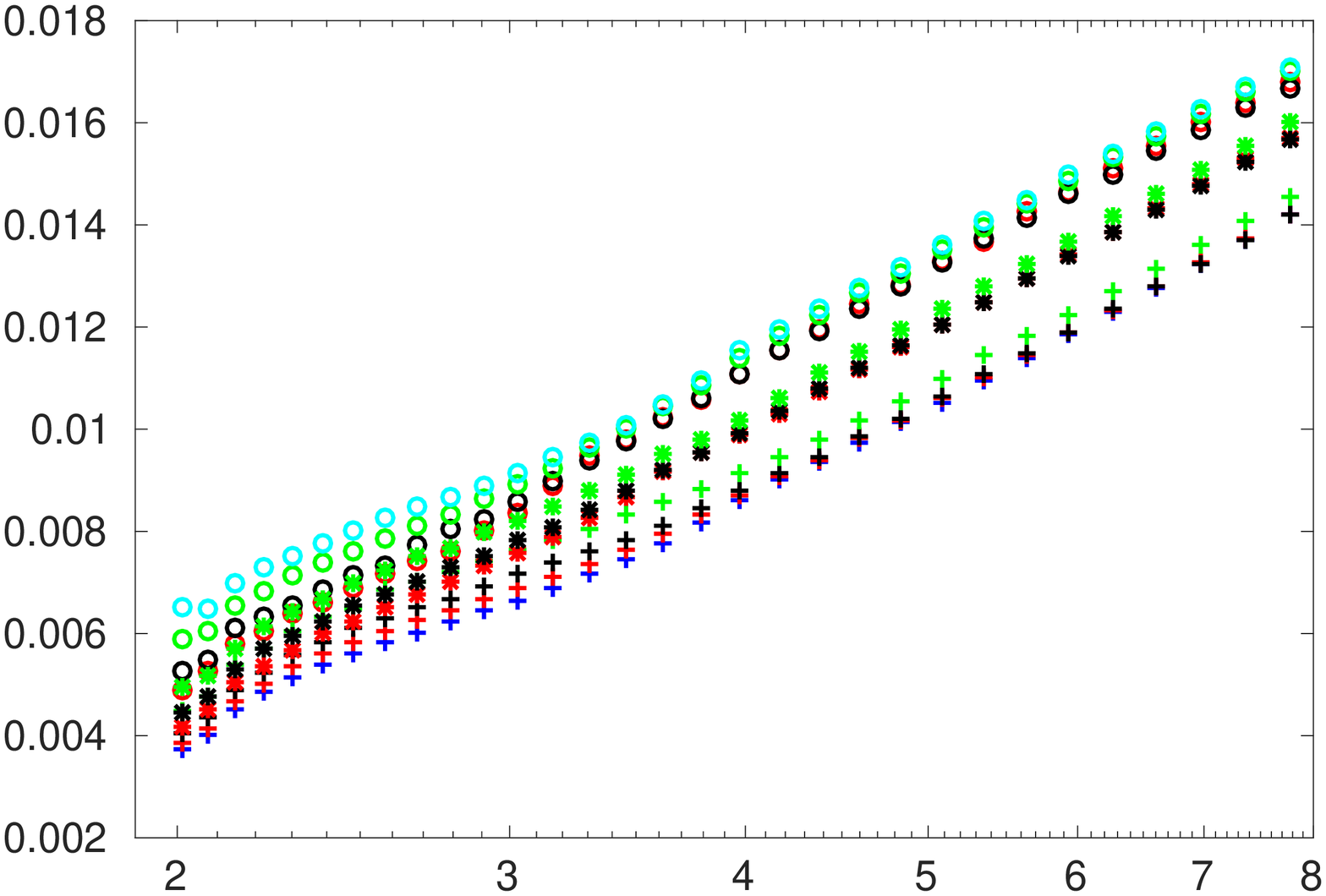}
      \put(-190,125){{\large $(c)$}} 
           \put(-195,70){{ $\kappa $}}
      \put(-95,0){{ $r/a$}}
   \caption{(a) Collision kernel at contact, (b) particle relative normal velocity and (c) Extended collision kernel of the different particle pairs as a function of the distance between the particle surfaces.}
   \label{fig:kern}
\end{center}
\end{figure}

The relative position between the particles in the flow determines part of the collision dynamics, but the full scenario becomes clear when considering  also the particle pair relative velocity in the normal direction.  The normal relative velocity between the particle pair as function of the distance $r$ reads, $dv_n(r)= (\textbf{u}_i-\textbf{u}_j). \frac{\textbf{r}_i - \textbf{r}_j}{|\textbf{r}_i - \textbf{r}_j|}$, where $\textbf{u}_i$ and $\textbf{r}_i$ are the velocity and position vector of particle $i$. 
The multiplication of the ensemble average of the negative part of the normal relative velocity, $dv^-_n(r) = dv_n(r) |_{<0}$, and the Radial Distribution Function gives the collision kernel: $\kappa (r)=   \langle dv^-_n(r)  \rangle \cdot  g(r)$. Note that the collision kernel is properly defined at the pair contact; however here we use the extended formula as a function of $r$ to understand how the approaching dynamics governs the collisions.

We display  the Radial Distribution Function (RDF) pertaining the different bi-disperse cases under investigation in figure \ref{fig:gr}(a). In this work, the particle-pair statistics are reported only for the region in the middle of the channel, in the range $0.25 < y/2h < 0.75$, in order to exclude wall effects.
We thus compare the particle dynamics in the homogenous region of the channel. {Note that here the average density of particle pairs in this region is used in the calculation of $n_0$.}
All the profiles are characterised by strong segregation at $r\simeq2a$ where $a$ is defined by the distance between the particle pairs at contact in each case; i.e. all the profiles start at $r=2a$ corresponding to the contact between the pairs. 
As $r$ increases $g(r)$ decreases and tends to unity when $r \gg 2a$ denoting a random distribution. As it can be seen in figure \ref{fig:gr}(b) where the Radial Distribution Function is shown at $r=2a$, increasing the percentage of large particles generally increases the segregation at contact for all possible pairs. We note that for the monodisperse cases, the RDF at contact is higher for larger particles. For bidisperse suspensions, the pairs formed by large-large and small-small particles show higher segregation than mixed pairs. Hence particles tend to cluster more with those of the same size, probably because their dynamics are more similar.

The collision kernels for particles at contact are depicted in figure \ref{fig:kern}(a). The trend is not the same as for the particle-pair segregation indicating the importance of the relative velocity for a collision to occur. In particular, we note that the normalized probability of collisions is the highest for pairs of large particles, followed by the probability of collisions between particles of different sizes and finally lowest for pairs of only small particles. Since the collision kernel is determined by the RDF times the average approaching velocity at contact, the observed differences are due to the pair dynamics. In figure \ref{fig:kern}(b) we therefore show the absolute value of the negative normal relative velocity as a function of the distance between particles. As expected, the absolute relative velocities increases monotonically with $r$, see also \cite{Lashgari16}; however the rate is different for different particle sizes. Large particles exhibit the highest normal relative velocities at all separation distances while the opposite is true for small particles. This result is not surprising since large particles have more inertia and are expected to collide even when starting at larger separations, and the dynamics at larger separation are dominated by the most energetic large-scale turbulent fluctuations. 
As mentioned above, the product of the relative approaching velocity and the RDF determines the collision kernel, whose scale-by-scale behavior is shown in figure \ref{fig:kern}(c). This quantity reduces decreasing inter-particle distances until it reaches the value at contact shown in the
panel(a). {Finally we note that the negative (approaching) relative velocity of pair-particles at $r \approx 2a$ is small but not negligible. 
The relative velocities are reduced by the strong lubrication force acting when the distance between the two particles is small. On the other hand, the probability of finding pair-particles at contact is high as shown in the plot of the radial distribution function. Thus, on average, a third particle approaching a pair feels the variation in the particle dynamics at $r=4a$. This explains the change in the slope of the relative velocity and collision kernel around $r=4a$.}

\section{Conclusion and discussion}

We study turbulent channel flow of binary mixtures of finite-size particles numerically. An Immersed Boundary Method is employed to simulate the motion of finite-size neutrally-buoyant spheres by ensuring that the no-slip and no-penetration boundary conditions are satisfied on the surface of each particle. Short-range interactions are modelled with an analytical lubrication force correction and a soft-sphere collision model.
The bulk Reynolds number and the total volume fraction of the solid phase are kept constant, $Re_b=5600$ and $\Phi=0.2$, while the ratio between the volume fraction of the small and large particles is varied. The ratio between the channel height and the diameter of the large and small particles are $2h/d_s=30$ and $2h/d_l=20$, respectively. Five different particulate cases are simulated, denoted as $0-100$, $25-75$, $50-50$, $75-25$ and $100-0$ according  to the percentage of volume fraction of small and large particles in the flow. In this study we have reported the bulk behaviour of the flow, wall-shear stress, turbulent statistics and particle dynamics for the different cases considered. 

The presence of the solid phase significantly alters the  fluid mean velocity with respect to that pertaining the unladen flow. In particular, the mean velocity profile is less blunt and the additive constant, $B$, of the log profile, $U^+= \frac{1}{\kappa} \mathrm{ln} y^+ + B$, decreases considerably. This  indicates drag enhancement, as shown in \cite{Picano15} for monodisperse particle suspensions. The particle mean velocity profile shows large slip with respect to the fluid phase in the near wall region and almost no-slip in the bulk of the channel. The slip at the wall increases with the relative percentage of large particles. Overall, the difference between the mean velocities of the binary mixtures and of the two mono-disperse flows at the same total volume fraction is small. {A similar behaviour has also been observed by \cite{Richter16} in simulations of dilute particulate turbulent flow with a two-way coupling model. Therefore, we expect that  the total particle volume fraction, and not the bidispersity of the particles, plays a major role in the bulk flow properties even at low particle concentrations.}

The profile of the {local volume fraction} is characterised by a homogenous particle distribution in the main body of the flow due to the turbulent mixing, while an inhomogeneous distribution is found near the wall where particle layering occurs. Considering the distribution of large and small particles separately, we observe a more intense layering of small particles. We remark that even in the case denoted $50-50$, i.e.\ with equal volume fraction of small and large particles, the layering appears essentially only in the small particle statistics, which therefore control the near wall dynamics. {Similar observations are made when considering the profile of the mean particle number density.}

We observe that the wall shear stress of the suspension parameterised by the friction Reynolds number does not change monotonically from the case of only small to only large particles, at same total volume fraction. The three cases $0-100$, $25-75$, $50-50$ share similar values, which are the lowest, while the case $100-0$ exhibits the highest value of  wall friction. Recently, a theoretical model has been proposed by \cite{Costa16} to predict the wall-shear stress and the mean velocity profile of turbulent suspensions of mono-disperse rigid particles. In the model two modifications are considered with respect to the unladen turbulent channel flow, the effective suspension viscosity and the formation of the particle layer at the wall, whose thickness scales with the particle size. Here, we apply the same model to estimate the friction Reynolds number of the binary mixtures. {To this end, we define an effective particle size responsible for the wall layering proportional to the distance from the wall of the local maximum of the volume fraction.  For all the cases under investigation, the prediction obtained with this estimate of the thickness of the particle layer shows a good agreement with the simulation data.} We finally remark that just using the effective viscosity the prediction is significantly underestimated, which confirms the key role of the particle layer on the drag increase. 

As concerns the  fluid and particle velocity fluctuations, we observe that the flow statistics vary monotonically between the two mono-disperse particulate cases. Comparing the statistics of particulate and unladen flows, we note that the turbulent activity reduces in the particulate cases when the particle stress becomes not negligible. The peak of the streamwise velocity fluctuations is shifted away from the wall due to the formation of the particle layer that hinders the production of turbulent kinetic energy. The opposite is true for the wall-normal component since the interactions between the particles and the wall increase the level of fluctuations close to the wall. As regards the statistics of the particle phase, we note that the fluctuations do not vanish at the wall; however similar to the fluid the particle statistics change monotonically between the two mono-disperse particulate cases.

We also study the particle dynamics, i.e.\ particle dispersion and collision  for the different particle-laden flows. 
We observe that the two mono-disperse cases show a similar rate of particle dispersion in the spanwise direction, whereas a non-monotonic behaviour characterizes the bi-disperse mixtures. This is explained  by the disturbances introduced by the presence of particles of different sizes. We also study the particle collision dynamics by computing the Radial Distribution Function and average normal relative velocity in the middle region of the channel, $0.25 \le y/2h \le 0.75$. These quantities are obtained separately for pairs consisting of two small, two large and one small and one large particle. The highest collision kernel is obtained for large-large particle pairs, followed by the mixed pairs and finally by the small-large particle pairs. This trend is  determined by the relative velocity of approaching particles that is highest for larger particles. Conversely, we note from the RDF that particles of the same size tend to be more clustered at contact. 

To conclude, we observe that the major effect of bi-dispersity is in the near-wall dynamics and in particular in the wall-layering. This aspect has macroscopic consequences such as the non-monotonic variation of the overall drag when changing the relative percentage of small and large particles.

\section*{Acknowledgement}
This work was supported by the European Research Council Grant no. ERC-2013-CoG-616186, TRITOS and by the Swedish Research Council Grant no. VR 2014-5001. The authors acknowledge computer time provided by SNIC (Swedish National Infrastructure for Computing) and the support from the COST Action MP1305: Flowing matter.

 \section*{Appendix A. Laminar flow of binary mixture of spheres}

In this Appendix, we discuss the results of simulations of binary mixtures  in laminar channel flow; i.e. we use the same setup and specifications used for the turbulent cases presented above and decrease the bulk Reynolds number to $Re_b=1000$. The simulations are initiated from the turbulent cases:  we observe the level of fluctuation to decrease in time and then to level off. Once the steady state laminar condition is reached, we compute the statistics presented here.  
 
\begin{figure}
\begin{center}
   \includegraphics[width=5.5cm]{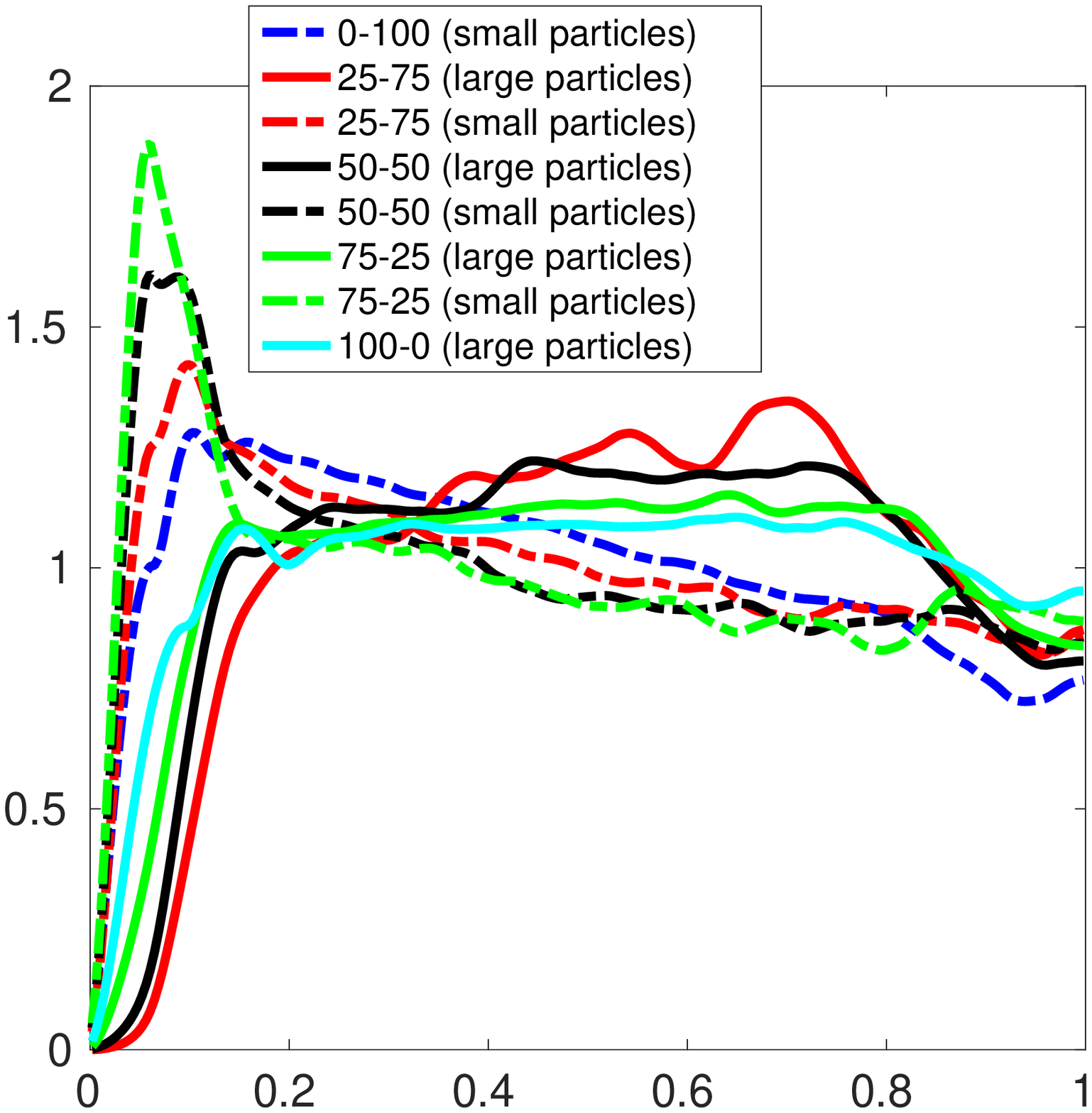}
     \put(-160,155){{\large $(a)$}} 
       \put(-170,80){{\large $\phi/\Phi$}} 
      \put(-80,0){{\large $y/h$}}
         \includegraphics[width=7cm]{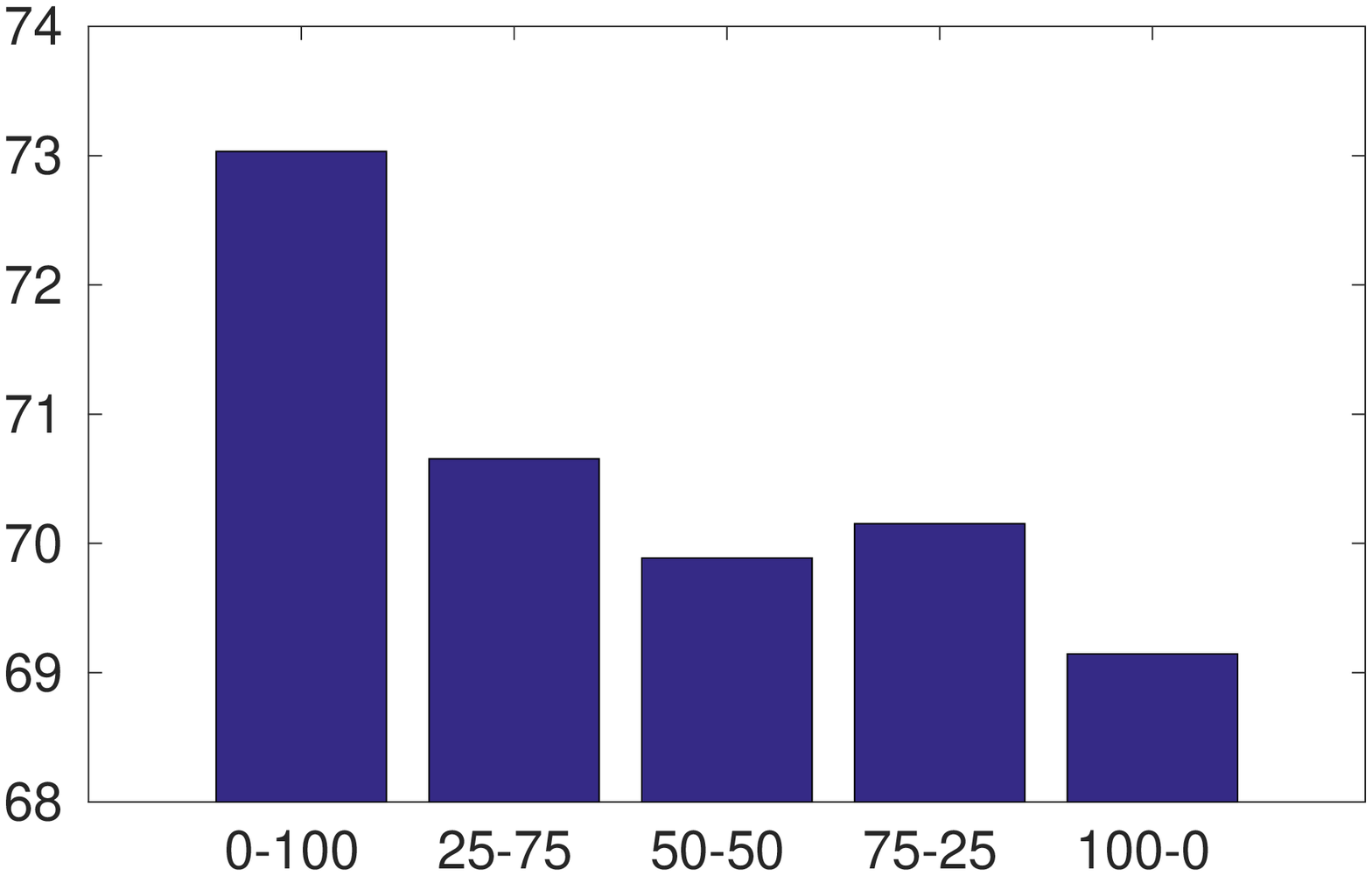}
        \put(-170,125){{\large $(b)$}} 
        \put(-15,60){{\large $Re_\tau \nu/\nu_e$}}
   \caption{(a) Profile of local particle concentration and (b) Friction Reynolds number for the five bi-disperse flows in laminar regime ($Re_b=1000$). In (a) the contribution of small and large particles are shown separately.
   }
   \label{fig:phi_lam}
\end{center}
\end{figure}

 {The wall-normal profiles of the local volume fraction for the five particulate cases at $Re_b=1000$ are shown in figure~\ref{fig:phi_lam}(a). Comparing this result with the one of the turbulent flows we observe a significantly stronger segregation of small particles close to the wall. The introduction of large particles in the suspension pushes the small particles away from the bulk of the flow towards the near wall region. Indeed the concentration of small particles is maximum at the wall and decreases monotonically towards the centre of the channel. The strong particle-particle and particle-wall interactions in the dense suspension move the position of the peak from (slightly closer to the wall in this case) the equilibrium of a single sphere in Poiseuille flow at the same Reynolds number \cite[][]{Segre61,Asmolov99}. As regards the concentration profile of the large particles, we also observe a second peak  in the region between the wall and the core of the channel. This peak is more evident for the case $25-75$, when the number of large particles is lower. This finding is inline with the experimental data of \cite{Matas04} where the formation of a second peak in the local volume fraction profile, close to the pipe centre, is reported for large particles. The formation of the second peak is attributed to the concavity of the lift force profile which is predicted by asymptotic theory for channel flow at large Reynolds numbers. Since we compute the statistics after the flow reaches statistically steady state, we believe the the formation of the second peak is not a transient phenomenon as it is questioned in the final discussion by \cite{Matas04}. }

In figure~\ref{fig:phi_lam}(b) we show the friction Reynolds number of the laminar cases normalised by the effective viscosity of the suspension. The normalised wall shear stress of the laminar cases are significantly lower than those of the turbulent flows. This is directly connected to the reduction of the fluid and particle Reynolds stresses in laminar condition. Unlike in the turbulent cases, the largest wall shear stress is obtained for the case  $0-100$, i.e., only small particles. This is attributed to the higher dissipation induced by small particles as these are characterized by largest surface area for the same volume fraction,  as indicated by \cite{Elghobashi94}.   

\bibliography{mybibfile}
\bibliographystyle{jfm}

\end{document}